\documentclass[onecolumn]{IEEEtran}
\usepackage[T1]{fontenc}
\usepackage[latin9]{inputenc}
\usepackage{array}
\usepackage{float}
\usepackage{amsmath}
\usepackage{amsthm}
\usepackage{amssymb}
\usepackage{graphicx}
\usepackage{setspace}
\usepackage[unicode=true,
 bookmarks=true,bookmarksnumbered=true,bookmarksopen=true,bookmarksopenlevel=1,
 breaklinks=false,pdfborder={0 0 0},pdfborderstyle={},backref=false,colorlinks=false]
 {hyperref}
\hypersetup{pdftitle={Your Title},
 pdfauthor={Your Name},
 pdfpagelayout=OneColumn, pdfnewwindow=true, pdfstartview=XYZ, plainpages=false}

\makeatletter

\providecommand{\tabularnewline}{\\}
\floatstyle{ruled}
\newfloat{algorithm}{tbp}{loa}
\providecommand{\algorithmname}{Algorithm}
\floatname{algorithm}{\protect\algorithmname}

\theoremstyle{plain}
\newtheorem{thm}{\protect\theoremname}[section]
\theoremstyle{definition}
\newtheorem{defn}[thm]{\protect\definitionname}
\theoremstyle{plain}
\newtheorem{lem}[thm]{\protect\lemmaname}
\theoremstyle{plain}
\newtheorem{prop}[thm]{\protect\propositionname}
\theoremstyle{definition}
\newtheorem{example}[thm]{\protect\examplename}
\theoremstyle{plain}
\newtheorem{cor}[thm]{\protect\corollaryname}

\usepackage[caption=false,font=footnotesize]{subfig}
\usepackage{cite}
\DeclareMathOperator*{\plim}{plim}

\makeatother

\providecommand{\corollaryname}{Corollary}
\providecommand{\definitionname}{Definition}
\providecommand{\examplename}{Example}
\providecommand{\lemmaname}{Lemma}
\providecommand{\propositionname}{Proposition}
\providecommand{\theoremname}{Theorem}

\begin{document}

\title{Asymptotically Optimal Resource Block Allocation With Limited Feedback}

\author{Ilai~Bistritz,~\IEEEmembership{Student Member,~IEEE} , Amir~Leshem,~\IEEEmembership{Senior Member,~IEEE}
\thanks{The authors are with the Faculty of Engineering, Bar-Ilan University,
Ramat-Gan, Israel, e-mail: ilaibist@gmail.com, leshema@biu.ac.il. 

This research was supported by the Israel Ministry of Science and
Technology under grant 3-13038, by a joint ISF-NRF research grant
number 2277/16 and by the ISF research grant under Grant 903/2013.

Parts of this paper were presented at the 2018 IEEE Wireless Communications
and Networking Conference \cite{bistritz2018efficient}.}}
\maketitle
\begin{abstract}
Consider a channel allocation problem over a frequency-selective channel.
There are $K$ channels (frequency bands) and $N$ users such that
$K=bN$ for some positive integer $b$. We want to allocate $b$ channels
(or resource blocks) to each user. Due to the nature of the frequency-selective
channel, each user considers some channels to be better than others.
The optimal solution to this resource allocation problem can be computed
using the Hungarian algorithm. However, this requires knowledge of
the numerical value of all the channel gains, which makes this approach
impractical for large networks. We suggest a suboptimal approach,
that only requires knowing what the $M$-best channels of each user
are. We find the minimal value of $M$ such that there exists an allocation
where all the $b$ channels each user gets are among his $M$-best.
This leads to feedback of significantly less than one bit per user
per channel. For a large class of fading distributions, including
Rayleigh, Rician, m-Nakagami and others, this suboptimal approach
leads to both an asymptotically (in $K$) optimal sum-rate and an
asymptotically optimal minimal rate. Our non-opportunistic approach
achieves (asymptotically) full multiuser diversity as well as optimal
fairness, by contrast to all other limited feedback algorithms.
\end{abstract}


\IEEEpeerreviewmaketitle{}

\section{Introduction}

The problem of multiple users accessing a common medium is a central
issue in every communication network \cite{Zhao2007,Hasan2013,Akkarajitsakul2011}.
The most common access schemes in practice are orthogonal transmission
techniques such as TDMA, FDMA , CDMA and OFDMA \cite{Seong2006,Zhao2015}.
Modern communication networks allocate each user a number of resource
blocks, each consisting of several OFDMA subcarriers that need not
be adjacent \cite{Ghosh2010a}.

When using a frequency division technique in a frequency-selective
channel, not all channels have the same quality. Furthermore, since
different users have different positions, the channel gains of users
in a given frequency band are independent. This introduces a resource
allocation problem between users and frequency bands (channels). Theoretically,
if the exact knowledge of all these channel gains were available,
an optimal solution for this resource allocation could be computed.
The performance gain that can be achieved by solving this problem
over a frequency allocation that ignores the channel gains is known
as multiuser diversity \cite{Yoo2005}. 

In FDD systems, the base station (BS) can only have knowledge of the
channel gains of each user if all users transmit this information
directly. In TDD systems, the BS can estimate the channel gain of
a given user from the reciprocity of the channel. However, the BS
must estimate the channel gains in all frequencies, not just the currently
used ones, so a wideband transmission is required from each user.
These transmissions need to be coordinated between users so that they
will not collide. Moreover, although the channels are indeed reciprocal,
the RF circuits are not, which calls for a calibration process \cite{Guey2004}. 

As networks grow and bandwidth expands, the number of channels increases
significantly. This is typical of OFDMA systems that tend to have
a large number of subcarriers. For example, in LTE \cite{Ghosh2010a}
there are from 128 to 2048 subcarriers in the downlink for each user.
This implies that reporting the channel state information (CSI) to
the BS inflicts a large communication overhead that makes such a scheme
infeasible. 

This paper deals with a channel allocation task with $K$ channels
where $N$ users simultaneously demand $b$ channels for each. This
occurs when users are backlogged; i.e., always have packets to transmit.
This represents a non-opportunistic approach where the BS waits for
a queue of $N$ users to form and only then computes a good allocation
for all the users in the queue. When users leave the network and others
replace them, the allocation is updated accordingly. 

We propose a novel limited feedback scheme that significantly reduces
feedback overhead compared to state of the art techniques \cite{Gesbert2004,Sanayei2007,Chen2006,Chen2008,Leinonen2009}.
The proposed technique provides a provably both asymptotically optimal
sum-rate \textbf{and} max-min fairness (which none of the current
methods provide). The proof applies to a wide range of fading distributions
that includes Rayleigh, Rician and m-Nakagami. Our technique uses
\textasciitilde 0.85 bits per user per channel for $K=N=20$, and
\textasciitilde 0.52 bits per user per channel for $K=512$ and $N=128$. 

In our approach, users only need to feed back the indices of their
$M$-best channels, and not their numerical values. If $M$ is large
enough, there is a high probability of an allocation of users to channels
where each user gets $b$ channels, all of which are from his $M$-best
channels. We prove that for $M\geq\left(b+1\right)\left(1+\varepsilon\right)\ln K$,
for $\varepsilon>0$, this probability approaches one as $K\rightarrow\infty$.
While our approach is asymptotic, the probability that such a perfect
matching allocation exists exceeds 99\% for $K=20$ channels and $N=20$
users. 

\begin{figure}[t]
~~~~~~~~~~~~~~~~~~~~~~~~~~~~~~~~~~~~~~~~~~~~~\includegraphics[width=7cm,height=5cm]{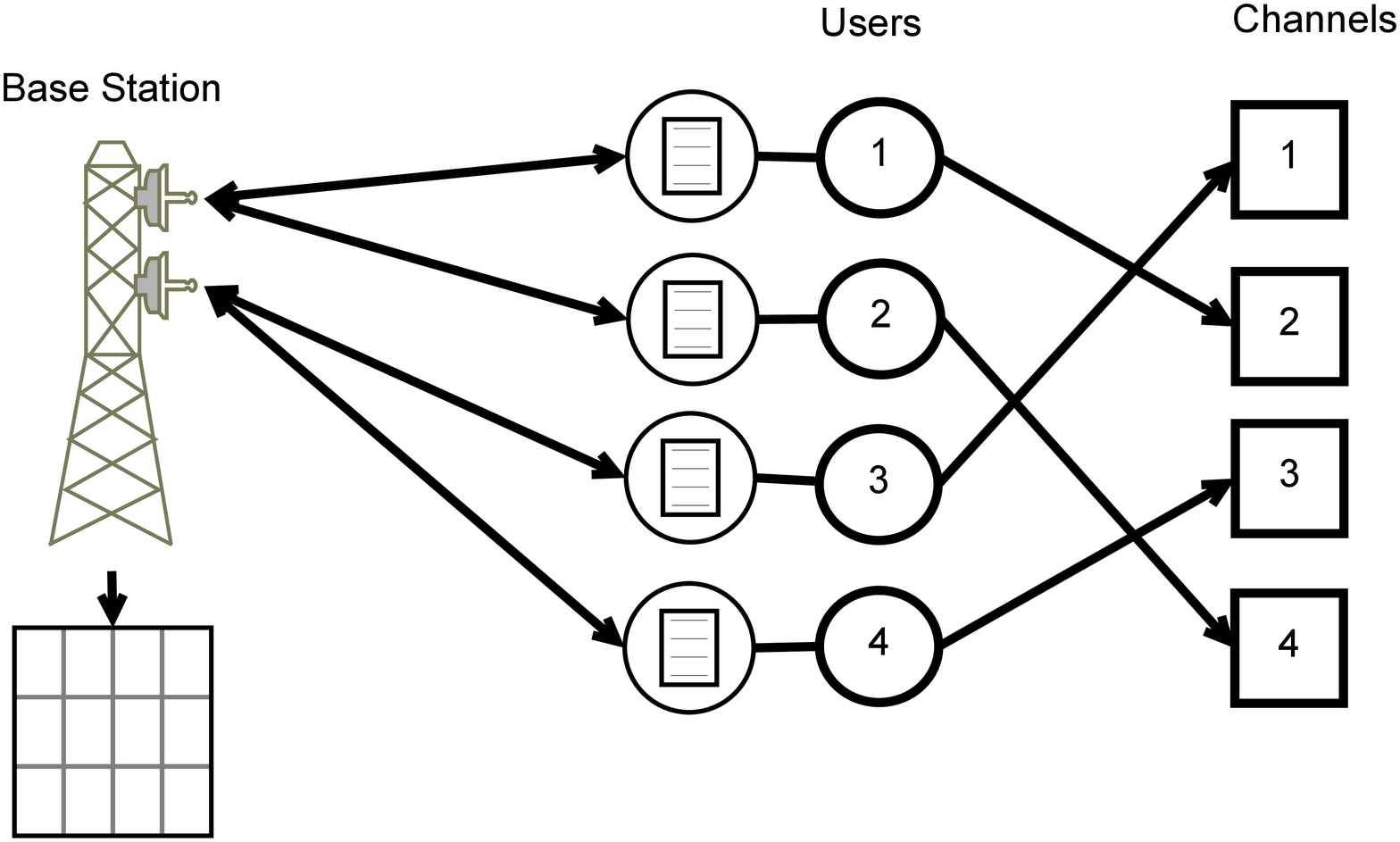}

\caption{\label{fig:The-System}The System}
\end{figure}

Figure \ref{fig:The-System} presents a toy example of our system
with $N=K=4$. Each user transmits the indices of his $M$ best channels
to the BS. The BS constructs the associated bipartite graph, computes
a perfect matching in this graph and transmits back the allocated
channel to each user. 

The existence of a perfect matching allocation relies on the theory
of random bipartite graphs. We define a novel random bipartite graph
where each user is represented by $b$ vertices (``agents'') in
the graph, all of which are connected to the $M$-best channels of
this user on the other side of the graph. We prove that this bipartite
graph has a perfect matching, with high probability. Due to its dependent
edges, our random bipartite graph is not an Erd\H{o}s--Rényi graph.
Consequently, proving the existence of a perfect matching requires
revisiting the original proof. Our proof is in the spirit of the original
result by Erd\H{o}s and Rényi for random bipartite graphs \cite{Erdos1964}.
A generalization of our results to the case of unequal numbers of
channels between users can be achieved using the same techniques.
It is omitted due to the complicated formalism. However, we present
detailed simulations of the CDF of the rates (not just the means)
that indicate that our results are also valid for an unequal number
of channels between users.

Throughout this paper, we use the term ``channel''. This may represent
several united subcarriers if the coherence bandwidth is large and
they are highly correlated (e.g., a resource block). In other words,
a ``channel'' is a designed unit that should optimally be the smallest
independent resource block in terms of the statistics of the corresponding
channel gain.

\subsection{Related Work}

There is extensive literature on OFDMA resource allocation (see the
surveys in \cite{Yaacoub2012,Sadr2009} and the references therein).
These allocations deal with the subcarrier assignment and the power
allocation over these subcarriers. The state of the art algorithms
achieve a close to optimal sum-rate with reduced computational complexity
compared to the optimal solution. In many studies, it has been shown
that an equal power allocation between the subcarriers of each user
achieves very close to optimal performance \cite{Hoo2004,Kim2005,Yu2006,Yu2002,Wong2008,Jang2003}.
Hence, the problem of subcarrier allocation appears to be much more
significant than that of power allocation. However, all these approaches
assume that users estimate their CSI perfectly and feed all of it
back to the BS. While this assumption may be reasonable for small
networks, it is highly infeasible for large networks with a large
number of subcarriers, such as LTE \cite{Ghosh2010a}. 

In order to overcome the need for the whole CSI to exploit the selectivity
of the channel, many works have suggested suboptimal schemes with
limited feedback. For an excellent overview of limited feedback systems,
see \cite{Love2008}. The state of the art algorithms consider an
opportunistic scheduling approach, where each available channel is
assigned to a user with a high instantaneous channel gain \cite{Gesbert2004,Sanayei2007,Al-Harthi2007,Chen2006}. 

In \cite{Gesbert2004,Sanayei2007,Chen2006} each user transmits one
bit per channel to indicate if the corresponding channel gain exceeds
a threshold, while in \cite{Al-Harthi2007} each user transmits the
highest-rate modulation scheme he can support out of a discrete set
of schemes (which requires more than one bit per channel). In \cite{Toufik2006,Svedman2007},
channels are grouped together into clusters and only the information
about the quality of each cluster is transmitted by the user. This
grouping approach limits the amount of multiuser diversity that can
be achieved, so that the reduced feedback has a performance cost.
This cost is avoided if the grouped subcarriers are highly correlated,
but in this case the multiuser diversity is smaller. We argue that
for highly correlated channels, using grouping cannot be considered
a real feedback reduction, since assigning two bits to two highly
correlated subcarriers is a waste of feedback to begin with. In this
work we use the term ``channel'' to refer to the smallest group
of subcarriers for which the \textquotedbl channels\textquotedbl{}
are independent, which can be considered a kind of grouping.

In \cite{Chen2008}, the grouping and thresholding approaches were
combined to achieve a feedback of less than one bit per user per channel,
which leads to a compromise of suboptimal performance (even asymptotically).
In \cite{Leinonen2009}, the $M$-best channels approach was used
for opportunistic scheduling. Users are arranged in a queue, and each
of them transmits the indices of his $M$-best channels to the BS.
The BS sequentially assigns each user the best channel from these
$M$ channels if it is available, and a random channel if none of
them are. The thresholding approach requires the BS or the users to
know the fading distribution to compute the optimal threshold. The
$M$-best channels approach does not require any such knowledge, which
makes it more robust. 

All these approaches are opportunistic in nature and as such suffer
from an inherent unfairness. Some of the methods have suggested obtaining
some degree of fairness by designing the scheduler \cite{Svedman2007,Sanayei2007,Chen2006,Gesbert2004},
at the cost of a reduced sum-rate. In order to get a comparable sum-rate
with a proportional fairness scheduler, each user needs to wait until
he is chosen as the ``opportunistic user''. Therefore, fairness
can only be maintained by introducing very large delays for users,
which is far from being desirable or practical. 

The rationale behind the opportunistic approach is to exploit the
multiuser diversity. By contrast, here we argue that this widespread
approach is too conservative and that achieving multiuser diversity
should not come at the expense of fairness. We argue that fairness
is undermined because all these approaches schedule a user for each
channel separately. Somewhat surprisingly, when analyzing all the
channels together, each user is very likely to be ``an opportunistic
choice'' for some available channel. This calls for a more sophisticated
mathematical analysis of the random matchings between users and channels
as we employ in this paper, in contrast to the simplistic arguments
dealing with either one user or one channel at a time.

In contrast to \cite{Chen2006,Chen2008,Leinonen2009,Sanayei2007},
our work is not limited to Rayleigh-fading channels, but applies to
a much broader class of exponentially-dominated tail distributions.
Besides Rayleigh fading, this class includes Rician, m-Nakagami and
others.

In \cite{Naparstek2013,Naparstek2014,Naparstek2014a}, a distributed
channel allocation for Ad-Hoc networks was introduced, based on a
carrier-sensing multiple access (CSMA) scheme and a thresholding approach
for the channel gains. Although considering a different scenario,
the analysis in \cite{Naparstek2013,Naparstek2014,Naparstek2014a}
also used a random bipartite graph, albeit a very simplified one.
The random bipartite graph in \cite{Naparstek2013,Naparstek2014,Naparstek2014a}
was an Erd\H{o}s--Rényi graph (i.e., existence of edges is independent),
for which perfect matching existence results are well known \cite{Erdos1964}.
There are no perfect matching existence results that can be employed
for our graph, so a novel analysis is required. Our random bipartite
graph differs from an Erd\H{o}s--Rényi graph in two key ways. First,
we use the $M$-best channels approach instead of the thresholding
approach. This means that each user node is connected to exactly $M$
channel nodes, and not only on average. Second, we allow for an allocation
of $b>1$ channels for each user, which leads to $b$ identical vertices.
Due to these two characteristics, the edges in our random bipartite
graph are not independent. Moreover, the analysis in \cite{Naparstek2013,Naparstek2014,Naparstek2014a}
was limited to bounded channel gains and only proved the asymptotic
optimality of the expected sum-rate. We prove asymptotic optimality
for the random sum-rate in probability. More importantly, we prove
asymptotic optimality for the random minimal rate in probability. 

The existence of perfect matching allocations is closely related to
the pure Nash equilibrium (NE) of interference games. In our previous
work \cite{Bistritz2015,Bistritz2018} we designed a game where all
pure NE are an almost (or an exact) perfect matching between users
and channels. We assumed that each user was allocated a single channel.
The results of this paper can be exploited to generalize our previous
results to the case of $b>1$ channels per user.

\subsection{Outline}

The remainder of this paper is organized as follows: In Section II
we formulate the channel allocation problem and present our limited
feedback scheme. In Section III we prove that as $K\rightarrow\infty$,
the probability that a perfect matching exists in the user-channel
graph approaches one for $M\geq\left(b+1\right)\left(1+\varepsilon\right)\ln K$
(Theorem \ref{thm:Main}) and that it cannot approach one for smaller
$M$ (Lemma \ref{M is not fixed}). \textbf{\textit{This perfect matching
is an allocation where each user gets $b$ of his $M$-best channels}}.
In Section IV we analyze the feedback requirements of our scheme and
propose an efficient encoding method. In Section V we prove that a
perfect matching allocation asymptotically attains \textbf{\textit{both
an optimal sum-rate and max-min fairness}} (Theorem \ref{thm:optimality}).
Section VI provides simulation results that support our analytical
findings, compare our algorithm to state of the art algorithms and
show that our results hold for a more general model. Section VII concludes
the paper.

\section{Problem Formulation}

In our model, there are $N$ users that are served by a BS that has
$K$ channels for allocation. Figure 1 depicts a toy example of our
system with $K=N=4$. We assume that $K=bN$ for some positive integer
$b$ and that each user is allocated $b$ channels. The parameter
$b$, which we treat as given, can be optimized beforehand. A generalization
of our analysis to a different number of channels per user is possible
using a more cumbersome proof. We later demonstrate in simulations
that our scheme works just as well in this case. Throughout the paper,
we use calligraphic capitals to denote sets (and events), bold capitals
to denote matrices and capitals or lower case letters to denote scalars.
Our assumptions on the channel gains are as follows:

{\textbf{{Channel Gains Assumptions}}}
\begin{enumerate}
\item For each user $n$, the channel gains $g_{n,1},...,g_{n,K}$ are $K$
i.i.d. random variables. Channel gains of different users are independent.
This is a widespread assumption in the channel allocation literature
\cite{Leinonen2009,Chen2006,Chen2008}. However, in contrast to \cite{Chen2006,Chen2008,Leinonen2009,Sanayei2007},
we do not assume Rayleigh fading or that the channel gains of different
users are identically distributed. 
\item Each user can perfectly measure each of his $K$ channel gains. This
widespread assumption \cite{Leinonen2009,Chen2006,Chen2008,Svedman2007,Sanayei2007,Toufik2006}
is necessary so that each user can evaluate his $M$-best channels.
In modern devices, multichannel sensing has become common practice
that can be easily achieved by a beacon sent from the BS or other
standard estimation techniques. Our approach does not require users
to transmit this entire information to the BS.
\item Channel gains are quasi-static, such that the coherence time is larger
than the time it requires to feed back the partial CSI (i.e., block-fading).
This is also a very common assumption \cite{Leinonen2009,Chen2006,Chen2008,Al-Harthi2007,Sanayei2007,Toufik2006,Gesbert2004},
since the multiuser diversity can only be exploited in scenarios where
this assumption holds.
\end{enumerate}
We assume that the transmission power is the same in each channel,
so no water-filling technique is used. This assumption is later justified
both analytically and in simulations that show that the water-filling
gain is negligible in our model for all practical SNR values. This
has been verified for many similar scenarios \cite{Hoo2004,Jang2003,Kim2005,Yu2002,Wong2008,Yu2006}. 

Denote by $\mathcal{A}_{n}$ the set of allocated channels of user
$n$. The achievable rates of the users in bits per channel use, $\left\{ R_{n}\right\} $,
are defined as
\begin{equation}
R_{n}\left(\mathcal{A}_{n}\right)=\sum_{k\in\mathcal{A}_{n}}\log_{2}\left(1+\frac{g_{n,k}^{2}P_{\max}}{bN_{0}}\right)\label{eq:1-2}
\end{equation}
where $P_{\max}$ is the maximal transmission power of each user and
$N_{0}$ is the variance of the Gaussian noise in his receiver. Often,
the goal of a channel allocation scheme is to maximize the sum-rate
of the users
\begin{equation}
\begin{array}{c}
\underset{\left(\mathcal{A}_{1},...,\mathcal{A}_{N}\right)}{\max}\sum_{n=1}^{N}R_{n}\left(\mathcal{A}_{n}\right)\\
s.t.\,\,\mathcal{A}_{n}\bigcap\mathcal{A}_{m}=\emptyset\,\,\forall n,m\,,n\neq m
\end{array}\label{eq:1}
\end{equation}
where the constraint means that each channel is allocated to a single
user (orthogonal transmission). However, maximizing the sum-rate often
comes at the expense of fairness.
\begin{equation}
\begin{array}{c}
\underset{\left(\mathcal{A}_{1},...,\mathcal{A}_{N}\right)}{\max}\underset{n}{\min}R_{n}\left(\mathcal{A}_{n}\right)\\
s.t.\,\,\mathcal{A}_{n}\bigcap\mathcal{A}_{m}=\emptyset\,\,\forall n,m\,,n\neq m
\end{array}\label{eq:3}
\end{equation}
Our goal is to design a limited feedback channel allocation scheme
that maximizes the sum-rate without compromising the minimal-rate
(fairness).
\begin{table}[t]
\caption{\label{tab:Notation-and-symbols}Notations and symbols used throughout
this paper.}

~~~~~~~~~~~~~~~~~~~~~~~~~~~~~~~~~~~~~~~%
\begin{tabular}{|c|>{\centering}p{6.5cm}|}
\hline 
$N$ & Number of users\tabularnewline
\hline 
$K$ & Number of channels\tabularnewline
\hline 
$b$ & Number of channels per user\tabularnewline
\hline 
$M$  & Number of best channels for each user\tabularnewline
\hline 
$g_{n,k}$ & Channel gain of user $n$ in channel $k$\tabularnewline
\hline 
$P_{\max}$ & Maximal transmission power\tabularnewline
\hline 
$X_{\left(i\right)}$ & The $i$-th smallest variable among $X_{1},...,X_{N}$\tabularnewline
\hline 
$\mathcal{E}_{K}$ & Probability that a PM does not exist with $K$ channels\tabularnewline
\hline 
\end{tabular}
\end{table}

\subsection{Perfect Matching Allocation Scheme}

In order for the BS to be able to compute the optimal solution of
\eqref{eq:1}, all users need to feed back all their channel gains
to the BS. Then, the optimal solution can be computed at the BS in
polynomial time using the Hungarian Algorithm \cite{Papadimitriou98}.
However, a feedback overhead of $K$ quantized numbers for each user
renders this approach impractical for cellular networks. Therefore,
suboptimal schemes with limited feedback overhead and good performance
are essential. 

In general, there is a fundamental tradeoff between the amount of
feedback used for the allocation and the performance achieved. By
choosing a specific feedback structure, one chooses a specific operating
point for this tradeoff. For example, by feeding back the values of
the $NK$ channel gains the optimal solution can be computed, and
by feeding back one bit per user per channel ($K$ bits per user),
state of the art thresholding schemes can be applied \cite{Gesbert2004,Sanayei2007,Chen2006}.

We propose a novel limited feedback scheme that requires less feedback
than current state of the art schemes. Not only that it does not perform
worse, it even performs much better. The scheme is summarized in Algorithm
\ref{alg:Scheme}. In our approach, each user transmits only the indices
of his $M$-best channels, without their numerical value. Feeding
back the indices of the $M$-best channels was proposed in \cite{Leinonen2009}
for an opportunistic channel allocation scheme. The novelty of our
approach is not in the feedback structure but in the fact that our
approach is non-opportunistic. This means that all channels and all
users are simultaneously considered in the computation of the allocation,
and not one by one. In particular, only in our scheme the results
of the next section about the optimal $M$ for a perfect matching
(PM) are valid. Since \cite{Leinonen2009} has nothing to do with
perfect matchings, it suffers, like any opportunistic method, from
an inherent non-fairness. We compare our performance to \cite{Leinonen2009}
in Section VI. 

The information available at the BS using this limited feedback can
be summarized using a bipartite graph defined as follows. Using this
information, the BS aims to compute an allocation where each user
gets $b$ of his $M$-best channels. Such an allocation does not necessarily
exist, and the existence probability depends on $M$. 
\begin{defn}
\label{def:A-user-channel-graph}A user-channel graph $\Gamma$ is
a balanced bipartite graph consisting of a user nodes set $\mathfrak{N}$
and a channel nodes set $\mathcal{\mathfrak{\mathcal{K}}}$. Every
user $n\in\mathcal{N}$ has $b$ user nodes (``agents'') and every
channel has a single channel node. An edge is connected between $n_{i}\in\mathfrak{N}$
and $k\in\mathcal{\mathfrak{\mathcal{K}}}$ if and only if channel
$k$ is one of the $M$-best channels for user $n$. A PM allocation
corresponds to a PM in this graph.
\end{defn}
\begin{algorithm}[H]
\caption{\label{alg:Scheme}Perfect Matching Limited Feedback Scheme}
\textbf{Setting} -There are $N$ users in a cell and $K=bN$ channels
for a positive integer $b$. Let $g_{n,k}$ be the channel gain of
user $n$ in channel $k$. Let $M=\left\lceil \left(1+b\right)\left(1+\varepsilon\right)\ln K\right\rceil $
for some $\varepsilon>0$.

\textbf{Periodically in the cell:}
\begin{enumerate}
\item The base-station (BS) broadcasts a common beacon on all $K$ channels. 
\item Each user $n$ independently -
\begin{enumerate}
\item Estimates the channel gains $g_{n,k}$ for each $k$.
\item Transmits the subset of his $M$-best channels to the BS, using $\log_{2}\binom{K}{M}$
bits (see Algorithm \ref{alg:Encoding}).
\end{enumerate}
\item Based on the information from 2(b), the BS constructs the bipartite
graph $\Gamma$ of Definition \ref{def:A-user-channel-graph}. 
\item The BS computes the maximal matching in $\Gamma$ (e.g., using the
algorithm of \cite{Naparstek2014a}). 
\item The BS transmits to each user the channels he is matched to according
to Step 4.
\begin{enumerate}
\item Theorem \ref{thm:Main} guarantees that with high probability this
matching will assign each user $b$ of his $M$-best channels (``perfect
matching'').
\item Theorem \ref{thm:optimality} guarantees that with high probability
this allocation will have asymptotically both optimal sum-rate and
max-min fairness.
\end{enumerate}
\end{enumerate}
\end{algorithm}

\begin{algorithm}[H]
\caption{\label{alg:Encoding}Feedback Encoding and Decoding}

\textbf{Encoding - }The subset $\mathcal{S}\subset\left\{ 1,...,K\right\} $
is mapped to the index $E\left(\mathcal{S},K\right)$, defined recursively
as
\begin{enumerate}
\item If $S=\emptyset$, then $E\left(\mathcal{S},K\right)=0$.
\item If $K$$\notin\mathcal{S}$, then $E\left(\mathcal{S},K\right)=E\left(\mathcal{S},K-1\right).$
\item If $K$$\in\mathcal{S}$, then $E\left(\mathcal{S},K\right)=\binom{K-1}{\left|\mathcal{S}\right|}+E\left(\mathcal{S}\setminus\left\{ K\right\} ,K-1\right).$
\end{enumerate}
\textbf{Decoding - }The index $e$ is mapped to the subset $D\left(e,K,M\right)$,
defined recursively as 
\begin{enumerate}
\item If $e=0$, then $D\left(0,K,M\right)=\emptyset$.
\item If $e<\binom{K-1}{M}$ then $D\left(e,K,M\right)=D\left(e,K-1,M\right)$.
\item If $e\geq\binom{K-1}{M}$ then $D\left(e,K,M\right)=D\left(e-\binom{K-1}{M},K-1,M-1\right)\bigcup\left\{ K\right\} $.
\end{enumerate}
\end{algorithm}

From the perspective of the theory of random bipartite graphs, our
graph is not typical. Each left side node has exactly $M$ edges whereas
each right side node may have any degree. Additionally, groups of
$b$ consecutive left side nodes are connected to the exact same right
side nodes. In particular, this means that edge existence is not independent
and that the graph is not an Erd\H{o}s--Rényi graph. This is why
a novel proof for the existence of a PM is needed.

\begin{figure}[t]
~~~~~~~~~~~~~~~~~~~~~~~~~~~~~~~~~~~~~~~~\includegraphics[clip,width=8cm,height=4.5cm]{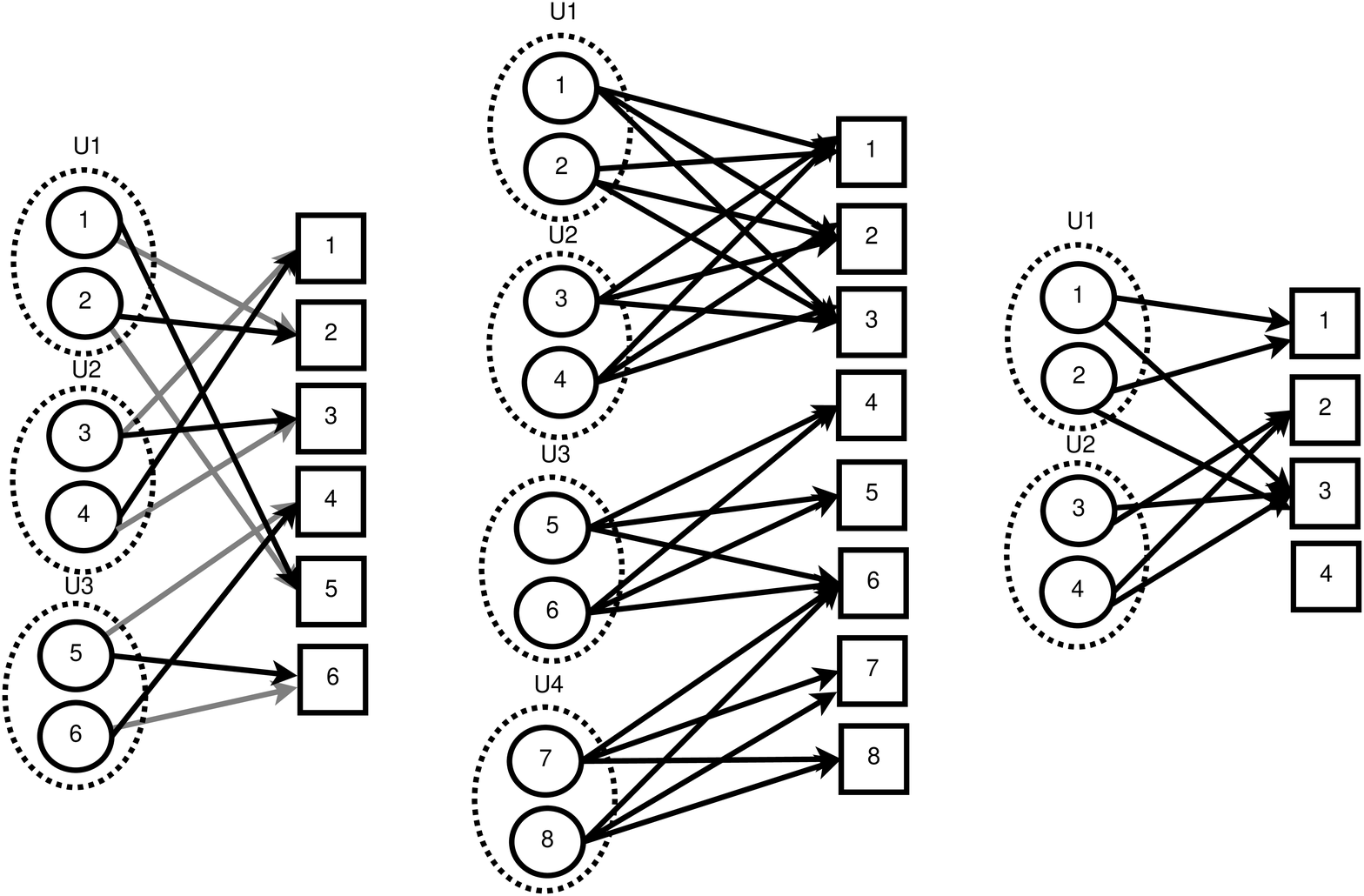}

\caption{\label{fig:Players-Channels-Graphs}User-channel graphs. }
\end{figure}
Examples of user-channel graphs can be seen in Figure \ref{fig:Players-Channels-Graphs}.
In the left graph $N=3$, $b=2$ and $M=2$. A PM exists that consists
of the grey edges. In the right most graph $N=2$, $b=2$ and $M=2$.
No user is connected to channel 4 so no PM can exist. In the central
graph, $N=4$, $b=2$ and $M=3$. Although each channel is connected
to at least two user nodes, a PM does not exist. No matter which two
channels are allocated to nodes 1 and 2, nodes 3 and 4 will be left
with only one desired channel for both of them. 

\section{Asymptotic Existence of PM Allocations}

Our goal is to show that a PM allocation, where each user only gets
$M$-best channels, exists. There is a trade-off for the parameter
$M$. A large value of $M$ increases the probability that a PM allocation
exists at the cost of reduced performance, since the $M$-best channel
gets worse as $M$ increases. The feedback communication overhead
also increases with $M$. Hence, it is important to identify the smallest
value of $M$ for which a PM exists. In this section we characterize
this optimal value of $M$ that is used in Algorithm \ref{alg:Scheme}. 

The main theorem of this paper is stated as follows:
\begin{thm}[Main Theorem]
\label{thm:Main} Let $K=bN$ for a positive integer $b$ that is
constant with respect to $N$. Let $\mathcal{E}_{K}$ be the event
that a PM, where each user gets $b$ channels that are among his $M$-best,
does not exist. If $M\geq\left(b+1\right)\left(1+\varepsilon\right)\ln K$
for some $\varepsilon>0$ then, for large enough $K$,
\begin{equation}
\Pr\left(\mathcal{E}_{K}\right)\leq\frac{1}{K^{\varepsilon+\frac{\varepsilon+1}{b}}}+\frac{\pi\sqrt{b}e^{3}}{12}\frac{1}{K^{\frac{3}{2}\varepsilon+\frac{\frac{3}{2}\varepsilon+1}{b}}}.\label{eq:4-1}
\end{equation}
Thus, a PM exists with a probability that approaches one as $K\rightarrow\infty$.
Furthermore, if $\varepsilon\geq1$, then the indicator of $\mathcal{E}_{K}$
converges to zero almost surely.
\end{thm}
\begin{IEEEproof}
See Appendix \ref{sec:Proofs}.
\end{IEEEproof}
For a PM to exist, every channel should be one of the $M$-best channels
for at least one user. The following lemma shows that for this to
occur, $M$ must increase with $K$ faster than $b\ln K$ and that
$M=b\left(1+\varepsilon\right)\ln K$ is enough for that purpose. 
\begin{lem}
\label{M is not fixed}Let $K=bN$ for a positive integer $b$ that
is constant with respect to $N$. Let $M>0$ and $\varepsilon>0$.
Let $\mathcal{E}_{0}$ be the event that there is a channel which
is not one of the $M$-best channels for any user.
\begin{enumerate}
\item If $M\leq b\ln K$ then $\underset{K\rightarrow\infty}{\lim}\Pr\left(\mathcal{E}_{0}\right)\geq\frac{1}{2}$.
\item If $M\geq b\left(1+\varepsilon\right)\ln K$ then $\underset{K\rightarrow\infty}{\lim}\Pr\left(\mathcal{E}_{0}\right)=0$.
\end{enumerate}
\end{lem}
\begin{IEEEproof}
See Appendix \ref{sec:Proofs}. 
\end{IEEEproof}

\section{Feedback Reduction}

The implication of Theorem \ref{thm:Main} is that Algorithm \ref{alg:Scheme}
requires of each user to feed back only the indices of his $M=\left\lceil \left(b+1\right)\left(1+\varepsilon\right)\ln K\right\rceil $
best channels. This requires $\log_{2}\binom{K}{M}$ bits from each
user, to encode a subset of size $M$ of a set of size $K$. Since
$\log_{2}\binom{K}{M}\leq M\log_{2}K$, the amount of feedback bits
per user per channel scales like $O\left(\frac{\log_{2}^{2}K}{K}\right)$,
which improves with the number of channels $K$, therefore making
our scheme especially appealing for large systems. In Algorithm \ref{alg:Encoding}
we present a practical method for encoding and decoding by indexing
the subsets that requires exactly $\log_{2}\binom{K}{M}$ bits, used
for the combinatorial number system \cite{Lehmer1964}. Its computational
complexity is $O\left(K\right)$.
\begin{prop}
The encoding of Algorithm \ref{alg:Encoding} requires $\log_{2}\binom{K}{M}$
bits.
\end{prop}
\begin{IEEEproof}
The fact that $E\left(\mathcal{S},K\right)$ is a bijection so each
subset has a unique index and vice versa is proven in \cite{Siddique}.
For each subset $\mathcal{S}\subset\left\{ 1,...,K\right\} $ such
that $\left|\mathcal{S}\right|=M$ we have
\begin{equation}
E\left(S,K\right)\leq\sum_{m=0}^{M}\binom{K-1-m}{M-m}\underset{\left(a\right)}{=}\binom{K}{M}\label{eq:5-1}
\end{equation}
where (a) follows from the hockey-stick identity \cite{Siddique}.
Hence, all the subsets $\mathcal{S}\subset\left\{ 1,...,K\right\} $
such that $\left|\mathcal{S}\right|=M$ are mapped by $E\left(\mathcal{S},K\right)$
to the indices 1 to $\binom{K}{M}$. 
\end{IEEEproof}
State of the art limited feedback schemes \cite{Gesbert2004,Sanayei2007,Chen2006}
use a threshold for each channel gain. This means that each user has
to send a packet of $K$ data bits in each coherence time of the channel
(typically several milliseconds). Already in LTE $K=2048$ is a possibility
\cite{Ghosh2010a}, which results in a feedback (uplink) data rate
of 409.6kbps for a coherence time of 5 milliseconds, more than four
times the bandwidth of a voice call. In order to maintain multiuser
diversity this feedback channel should be active very often. Besides
creating a significant communication overhead in the uplink, this
might not even be feasible due to battery constraints. Since the number
of channels will increase in 5G systems, reducing the feedback amount
is necessary if channel allocation with good performance is to be
employed. 

In general, in the high-SNR regime, the capacity of each user scales
like $bB_{W}\ln\left(\ln N\right)$, where $B_{W}$ is the bandwidth
of a channel and $\ln\left(\ln N\right)$ is from the multiuser diversity
in Rayleigh fading channels \cite{Chen2008}. For a thresholding method
the feedback rate scales like $bN$, so the feedback becomes even
larger than the data rate in large systems. In \cite{Chen2008}, it
was suggested to mitigate this overhead by grouping channels, at the
cost of a reduced sum-rate (and minimal rate). Our method requires
a feedback rate that scales like $\ln^{2}N$ instead of like $N$
as the thresholding schemes. Therefore, for modern systems with a
relatively large $B_{W}$, the feedback overhead of our scheme remains
negligible even for very large values of $N$. Hence, our method achieves
the necessary feedback reduction, not only without reducing the performance,
but actually improving it. While the sum-rate improvement is moderate,
the minimal rate is dramatically increased. 

\begin{figure}[t]
\begin{centering}
\textsf{\includegraphics[width=5cm,height=5cm]{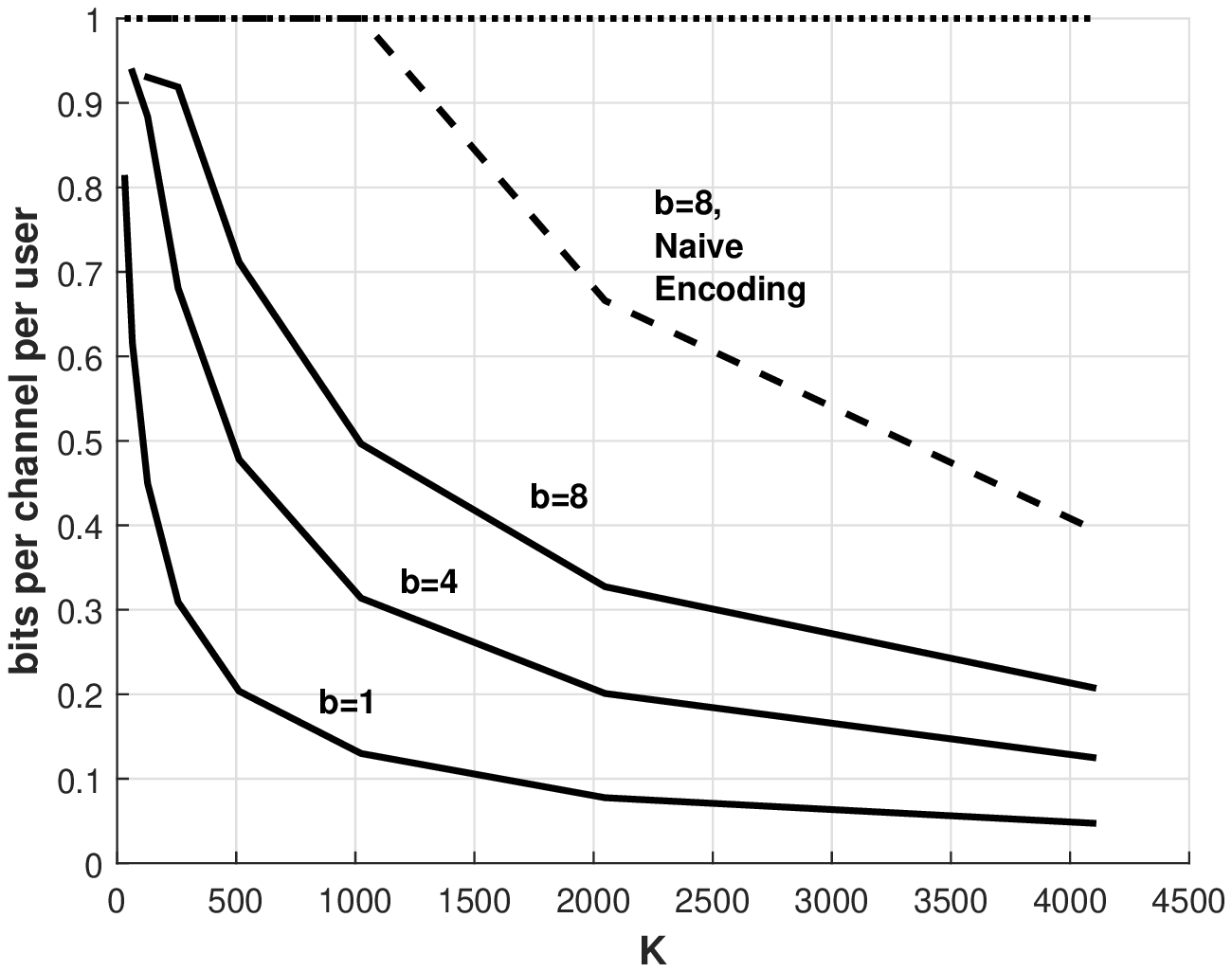}\includegraphics[width=5cm,height=5cm]{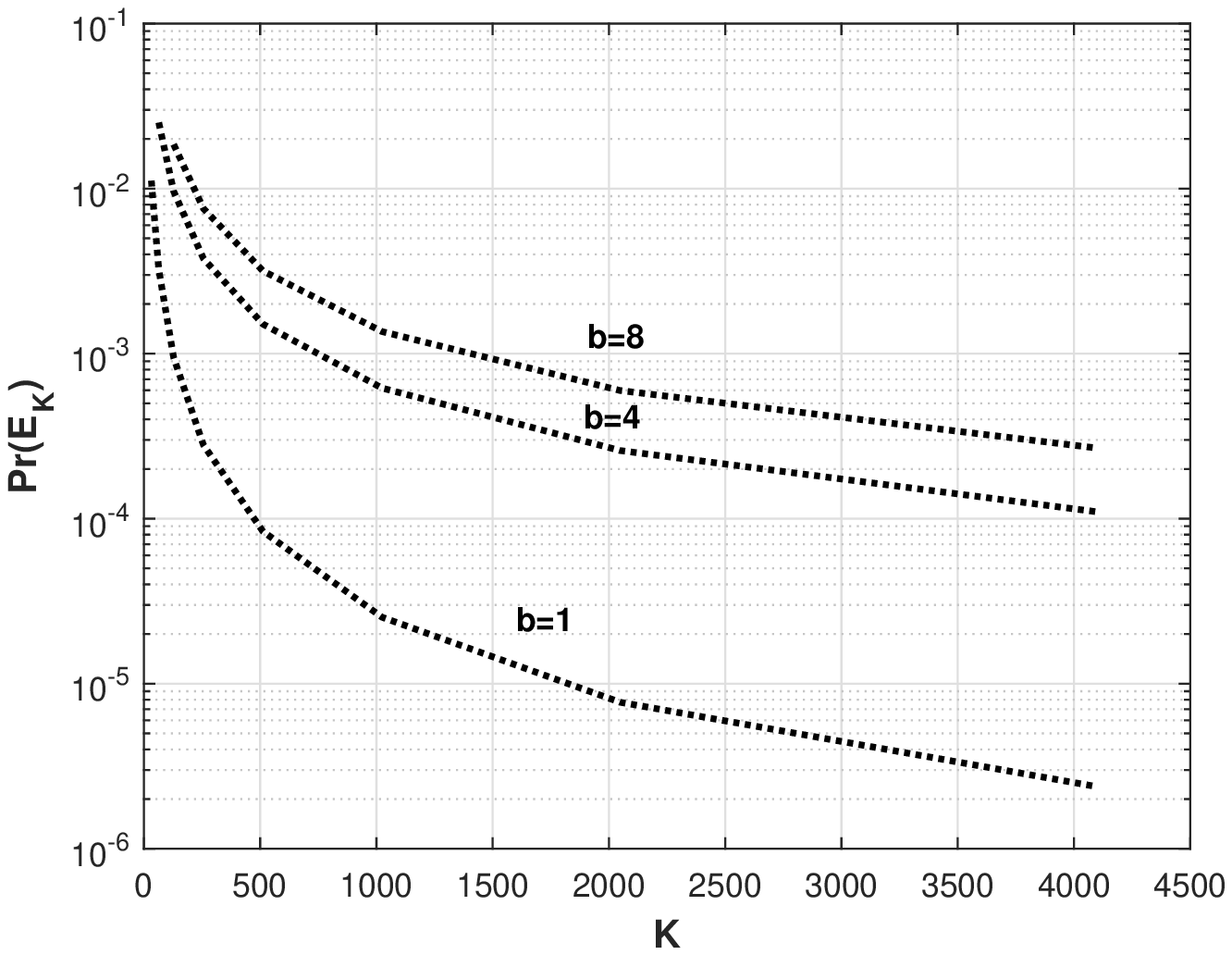}}
\par\end{centering}
\caption{\label{fig:FeedbackFigure}Number of bits for feedback in Algorithm
\ref{alg:Scheme}, and the corresponding upper bound for the non-perfect
matching probability.}
\end{figure}

In Figure \ref{fig:FeedbackFigure} we present the number of feedback
bits our algorithm requires as a function of the number of channels
$K$, for $b=1,4,8$ and $\varepsilon=0.3,0.7,0.8$, respectively.
The values of $\varepsilon$ are chosen in order to minimize the number
of feedback bits while keeping the probability that a PM does not
exist sufficiently small. We also show the number of bits required
by a naive encoding scheme that directly transmits the $M$ indices
using $M\log_{2}K=\log_{2}K\left\lceil \left(b+1\right)\left(1+\varepsilon\right)\ln K\right\rceil $
bits. It is evident that our algorithm always requires less than the
1 bits per user per channel that state of the art thresholding approaches
require \cite{Gesbert2004,Sanayei2007,Chen2006}. In comparison, the
optimal solution, given by the Hungarian algorithm, requires a quantization
of $q$ bits per user per channel, where a reasonable $q$ that causes
only a negligible performance degradation is $q=8$. For $N=256$
and $K=1024$, our approach requires less than 0.29 bits per user
per channel - a saving of more than 70\% in feedback overhead. For
$K=2048$ this becomes a saving of more than 80\%. Even for small
values like $N=K=32$, our approach requires less than 0.8 bits per
user per channel. We also present the upper bound of Theorem \ref{thm:Main}
on the probability that a PM does not exist, which is always less
than 2\% and improves very fast as $K$ grows. Note that since the
exponents of the bound in \eqref{eq:4-1} are $\varepsilon+\frac{\varepsilon+1}{b}$
and $\frac{3}{2}\varepsilon+\frac{\frac{3}{2}\varepsilon+1}{b}$,
the bound is much lower for $b=1$ and starts to behave like $\frac{1}{K^{\varepsilon}}$
for large $b$ values. 

\section{Asymptotic optimality of PM Allocations}

In the previous section we  showed that $M=\left(b+1\right)\left(1+\varepsilon\right)\ln K$,
for some $\varepsilon>0$, is enough to guarantee that a PM allocation
exists. In this section we show how good this PM allocation actually
is. This has to do with the extent to which the $M$-best channel
of each user is worse than his best channel. This definitely depends
on the distribution of the channel gains. Luckily, all of the fading
distributions used in practice tend to belong to the following class:
\begin{defn}[Exponentially-Dominated Tail Distribution]
 \label{Exponentially-Dominated Tail}Let $X$ be a random variable
with a continuous CDF $F_{X}$. We say that $X$ has an exponentially-dominated
tail distribution if there exist $\alpha>0,\beta\in\mathbb{R},\lambda>0,\gamma>0$
such that
\begin{equation}
\underset{x\rightarrow\infty}{\lim}\frac{1-F_{X}\left(x\right)}{\alpha x^{\beta}e^{-\lambda x^{\gamma}}}=1.\label{eq:20}
\end{equation}
The following simple proposition validates that many commonly used
fading distributions have an exponentially-dominated tail. 
\end{defn}
\begin{prop}
\label{prop:The-Rayleigh,--Nakagami,}The Rayleigh, $m$-Nakagami,
Rice and Normal distributions are exponentially-dominated tail distributions.
\end{prop}
\begin{IEEEproof}
See Appendix \ref{sec:Proofs}.
\end{IEEEproof}
The fact that each user gets $b$ of his $M$-best channels, and that
$M$ is kept relatively small for exponentially-dominated fadings,
provides both sum-rate and fairness guarantees. This is a special
property of our allocation. In general, maximizing the sum-rate does
not provide any fairness guarantees for the users, as demonstrated
in the following example.
\begin{example}
Consider the following $K\times K$ rates matrix for the case of $b=1$,
where $u_{n,k}$ is the rate of user $n$ in channel $k$. Let $\delta>0$
be a small number. \renewcommand{\arraystretch}{0.5} 
\[
\boldsymbol{U}=\left(\begin{array}{ccccc}
\delta & 1 & 1 & \cdots & 1\\
1 & 2 & 1 & \cdots & 1\\
\delta & 1 & 1 & \cdots & 1\\
\vdots & \vdots & \vdots & \ddots & \vdots\\
\delta & 1 & 1 & \cdots & 1
\end{array}\right)
\]
For this case, the Hungarian algorithm assigns channel $k$ to user
$k$ for each $k$, resulting in a sum-rate of $K+\delta$ and a minimal
rate of $\delta$. Our PM approach assigns each user one of his $M$-best
channels. If $2\leq M\leq K-1$, our algorithm results in a sum-rate
of $K$ and a minimal rate of $1$. For a small enough $\delta$,
we obtain only a slight reduction in the sum-rate while having a dramatically
better minimal rate. 
\end{example}
The following theorem establishes the asymptotic optimality (in ratio)
of a PM allocation both in the sum-rate and minimal rate senses, for
exponentially-dominated tail fadings. 
\begin{thm}[Sum Rate and Fairness Optimality]
\label{thm:optimality}Let the channel gains be the i.i.d. variables
$g_{n,1},...,g_{n,K}$ , for each $n$. If, for each $n$ and $k$,
$g_{n,k}$ has an exponentially-dominated tail and $M=\left(b+1\right)(1+\varepsilon)\ln K$
for some $\varepsilon>0$, then 
\begin{equation}
\plim_{K\to\infty}\frac{\underset{\left(\mathcal{A}_{1},...,\mathcal{A}_{N}\right)\in\mathcal{P}}{\min}\sum_{n=1}^{N}R_{n}\left(\mathcal{A}_{n}\right)}{\underset{\left(\mathcal{A}_{1},...,\mathcal{A}_{N}\right)}{\max}\sum_{n=1}^{N}R_{n}\left(\mathcal{A}_{n}\right)}=1\label{eq:21}
\end{equation}
where $\mathcal{P}$ is the set of PMs in the user-channel graph,
and \textup{plim} denotes convergence in probability. Furthermore,
\begin{equation}
\plim_{K\to\infty}\frac{\underset{\left(\mathcal{A}_{1},...,\mathcal{A}_{N}\right)\in\mathcal{P}}{\min}\underset{n}{\min}R_{n}\left(\mathcal{A}_{n}\right)}{\underset{\left(\mathcal{A}_{1},...,\mathcal{A}_{N}\right)}{\max}\underset{n}{\max}R_{n}\left(\mathcal{A}_{n}\right)}=1.\label{eq:22}
\end{equation}
\end{thm}
\begin{IEEEproof}
Denote by $g_{n,\left(i\right)}$ the $i$-th smallest channel gain
among $g_{n,1},...,g_{n,K}$. Denote $\Phi=\left(\mathcal{A}_{1},...,\mathcal{A}_{N}\right)$.
We have
\begin{multline}
1\geq\frac{\underset{\Phi\in\mathcal{P}}{\min}\sum_{n=1}^{N}R_{n}\left(\mathcal{A}_{n}\right)}{\underset{\Phi}{\max}\sum_{n=1}^{N}R_{n}\left(\mathcal{A}_{n}\right)}\underset{\left(a\right)}{\geq}\frac{\underset{\Phi\in\mathcal{P}}{\min}N\underset{n}{\min}R_{n}\left(\mathcal{A}_{n}\right)}{\underset{\Phi}{\max}N\underset{n}{\max}R_{n}\left(\mathcal{A}_{n}\right)}\underset{\left(b\right)}{\geq}\underbrace{\frac{\underset{n}{\min}b\log_{2}\left(1+\frac{g_{n,\left(K-M+1\right)}^{2}P_{\max}}{bN_{0}}\right)}{\underset{n}{\max}b\log_{2}\left(1+\frac{g_{n,\left(K\right)}^{2}P_{\max}}{bN_{0}}\right)}}_{A_{K}}\\
\geq\frac{\log_{2}\left(\left(1+\frac{\underset{n}{\max}g_{n,\left(K\right)}^{2}P_{\max}}{bN_{0}}\right)\frac{\underset{n}{\min}g_{n,\left(K-M+1\right)}^{2}}{\underset{n}{\max}g_{n,\left(K\right)}^{2}}\right)}{\log_{2}\left(1+\frac{\underset{n}{\max}g_{n,\left(K\right)}^{2}P_{\max}}{bN_{0}}\right)}=1+\underbrace{\frac{2\log_{2}\left(\frac{\underset{n}{\min}g_{n,\left(K-M+1\right)}}{\underset{n}{\max}g_{n,\left(K\right)}}\right)}{\log_{2}\left(1+\frac{\underset{n}{\max}g_{n,\left(K\right)}^{2}P_{\max}}{bN_{0}}\right)}}_{B_{K}}\label{eq:24}
\end{multline}
where (a) follows since for every allocation, the sum-rate is larger
than $N\underset{n}{\min}R_{n}\left(\mathcal{A}_{n}\right)$ and smaller
than $N\underset{n}{\max}R_{n}\left(\mathcal{A}_{n}\right)$. Inequality
(b) follows from the definition in \eqref{eq:1-2} since each of the
$b$ channel gains user $n$ gets is at least $\underset{n}{\min}\,g_{n,\left(K-M+1\right)}$
in all $\left(\mathcal{A}_{1},...,\mathcal{A}_{N}\right)\in\mathcal{P}$
and at most $\underset{n}{\max}\,g_{n,\left(K\right)}$ in all allocations.
First note that for $g_{n,k}$ with a bounded distribution we have
$\plim_{K\to\infty}\frac{\underset{n}{\min}g_{n,\left(K-M+1\right)}}{\underset{n}{\max}g_{n,\left(K\right)}}=1$
since both the numerator and the denominator approach $F_{g}^{-1}\left(1\right)<\infty$,
so $\plim_{K\to\infty}B_{K}=0$. Theorem 7 in \cite{Bistritz2018}
uses an argument similar to \eqref{eq:24} to show that $\plim_{K\to\infty}A_{K}=1$
even if $g_{n,1},...,g_{n,K}$ are unbounded but have an exponentially-dominated
tail, based on showing that $\plim_{K\to\infty}\frac{\underset{n}{\min}g_{n,\left(K-M+1\right)}}{\underset{n}{\max}g_{n,\left(K\right)}}>0$
for exponentially-dominated tail distributions. Therefore \eqref{eq:21},\eqref{eq:22}
are obtained from the Sandwich Theorem.
\end{IEEEproof}
Theorem \ref{thm:optimality} has a stronger argument than simply
asymptotic optimality of the sum-rate and the minimal rate. Observe
that the denominator in \eqref{eq:22} consists of the best rate a
user can receive in any allocation, even an allocation that only favors
him. This rate is achieved when user $n$ is allocated his best $b$
channels from out of the entire $K$ channels. Hence, even the user
with the minimal rate, in all possible PM allocations, achieves asymptotically
full multiuser diversity. In fact, this is also true for the $bN$
agents as well, which leads to the following corollary.
\begin{cor}
\label{cor:WF}Let $K=bN$ for a positive integer $b$ that is constant
with respect to $N$. The ratio between the rate achieved by water-filling
to that of an equal power allocation converges in probability to one
as $K\rightarrow\infty$.
\end{cor}
\begin{IEEEproof}
Denote by $\left\{ P_{n,k}^{*}\right\} $ the set of the optimal transmission
powers given $\left\{ g_{n,k}\right\} $. If $g_{n,k}$ is increased
to $g_{n,\left(K\right)}$ for each $k$ and the transmission powers
are kept as $P_{n,k}^{*}$ for each $k$, then the sum-rate increases.
If for the increased channel gains the optimal power allocation is
used instead of $\left\{ P_{n,k}^{*}\right\} ,$ the sum-rate will
further increase. The optimal power allocation for identical channel
gains of $g_{n,\left(K\right)}$ is an equal power allocation. Using
this argument, we obtain 
\begin{equation}
\frac{\sum_{k\in\mathcal{A}_{n}}\log_{2}\left(1+\frac{g_{n,k}^{2}P_{\max}}{bN_{0}}\right)}{\sum_{k\in\mathcal{A}_{n}}\log_{2}\left(1+\frac{g_{n,k}^{2}P_{n,k}^{*}}{N_{0}}\right)}\underset{(a)}{\geq}\frac{\sum_{k\in\mathcal{A}_{n}}\log_{2}\left(1+\frac{g_{n,\left(K-M+1\right)}^{2}P_{\max}}{bN_{0}}\right)}{\sum_{k\in\mathcal{A}_{n}}\log_{2}\left(1+\frac{g_{n,\left(K\right)}^{2}P_{\max}}{bN_{0}}\right)}\geq\frac{\underset{n}{\min}\log_{2}\left(1+\frac{g_{n,\left(K-M+1\right)}^{2}P_{\max}}{bN_{0}}\right)}{\underset{n}{\max}\log_{2}\left(1+\frac{g_{n,\left(K\right)}^{2}P_{\max}}{bN_{0}}\right)}.\label{eq:25}
\end{equation}
In (a) we also used $g_{n,k}\geq g_{n,\left(K-M+1\right)}$ for each
$k$, which holds in a PM allocation with parameter $M$. 
\end{IEEEproof}

\section{Simulation Results}

In this section, we demonstrate our analytical results using numerical
simulations. We also validate the fact that our findings hold in a
broader model that includes correlated channels and an allocation
of unequal numbers of channels to users. All the rates are measured
in bits per second, assuming that the bandwidth of a channel is 15kHz
(as in LTE \cite{LTE2009}). We used a Rayleigh fading network; i.e.,$\left\{ g_{n,k}\right\} $
are Rayleigh distributed and $\left\{ g_{n,k}^{2}\right\} $ are exponentially
distributed. In each experiment, we ran 100 random realizations of
the channel gains. Unless otherwise stated, the transmission powers
were chosen such that the mean SNR for each link was 20dB. We used
the Hungarian algorithm \cite{Papadimitriou98} to compute the optimal
sum-rate solution. We computed the maximal matching in the bipartite
graph (which Theorem 2 guarantees is a PM) using the algorithm in
\cite{Naparstek2013}. This algorithm requires a time complexity of
$O\left(K\log_{2}^{2}K\right)$ instead of $O\left(K^{3}\right)$
for the Hungarian algorithm, which also could have been used for computing
a maximal matching (on the binary preferences matrix).

\subsection{Uncorrelated Identically-Distributed Channels}

In this subsection, the channel gains were uncorrelated with parameter
$\lambda=1$ for all exponential variables. We used the parameter
$M=\left\lceil 1.5\left(b+1\right)\ln\left(K\right)\right\rceil $
and ran the simulations for $N=10,25,50,75,100$. We found that for
every realization a PM existed. This implies that our asymptotic results
are valid for as low as $N=10$.

\begin{figure}[t]
\begin{centering}
\textsf{\includegraphics[width=7cm,height=5cm]{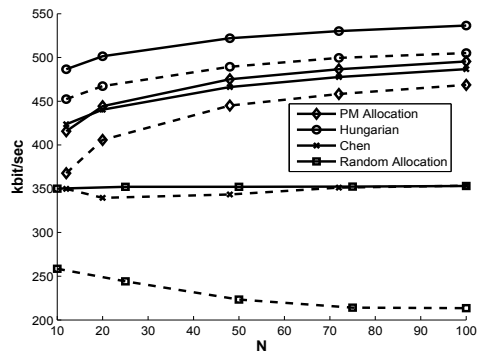}}
\par\end{centering}
\caption{\label{fig:The-sum-rate-of}Achievable rates as a function of $N$
for uncorrelated channels. Solid lines represent the mean rates and
dashed lines the minimal rates.}
\end{figure}

In Figure \ref{fig:The-sum-rate-of}, the mean and minimal rates are
presented as a function of $N$ for $b=4$. It is evident that the
performance of the PM allocation was already above 90\% of the optimal
allocation for $N=50$. As anticipated by our results, it improved
with $N$. We can also see that the minimal rate of the optimal allocation
was good, which was to be expected. The chance of a single user getting
$b=4$ bad channels is smaller than in the case of $b=1$. The random
allocation rates, which can be thought of as the result of an allocation
that ignores the CSI, were far behind. This effect is crucial especially
for the minimal rate. In \cite{Chen2006}, the BS assigns each channel
to one of the users that exceeded a certain threshold (possibly with
defined priorities between users). In order to make the algorithm
competitive in the minimal rate sense as well, each time multiple
users exceeded the threshold, we allocated the channel to the user
with the minimal number of channels thus far. We used $b=4$, so $K=4N$.
Our mean rate is slightly better than that of \cite{Chen2006} for
all $N\geq20$. The minimal rate of \cite{Chen2006} is much lower
(\textasciitilde 78\% of our minimal rate), and does not increase
with $N$ while our minimal rate does. Last but not least, we require
$\frac{1}{K}\log_{2}\binom{K}{M}$ bits per user per channel (0.81
for $N=10$ and 0.52 for $N=200$), while \cite{Chen2006} always
requires one bit per user per channel (exceeded the threshold or not).
Hence, our algorithm outperforms \cite{Chen2006} both in terms of
performance and feedback communication overhead.

\begin{figure}[t]
\begin{centering}
\textsf{\includegraphics[width=7cm,height=5cm]{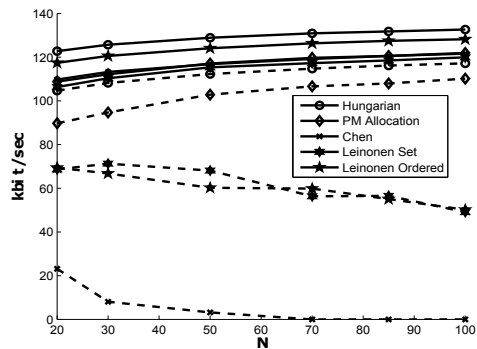}}
\par\end{centering}
\caption{\label{fig:ComparisonRatesb1}Achievable rates as a function of $N$
compared to Leinonen's Algorithm, Chen's Algorithm and the Hungarian
Algorithm. Solid lines represent the mean rates and dashed lines the
minimal rates.}
\end{figure}

In Figure \ref{fig:ComparisonRatesb1}, we compare our algorithm also
to that of \cite{Leinonen2009}, which has two versions. This algorithm
assigns a single channel to each user, so we had to choose $b=1$.
We used the same $M=\left\lceil 3\ln\left(K\right)\right\rceil $
for our algorithm and that of \cite{Leinonen2009}. The mean rate
of the set-best version is very similar to ours, with exactly the
same feedback communication overhead. The mean-rate in the ordered-best
version is a little better than our mean-rate, but has a significant
feedback cost. For $N=K=20$, the ordered-best requires \textasciitilde 2.25
bits per user per channel, while the set-best (and our approach) requires
\textasciitilde 0.87 bits per user per channel. For $N=K=100$, the
ordered-best requires \textasciitilde 0.98 bits per user per channel,
while the set-best and our approach require \textasciitilde 0.55
bits per user per channel. However, our major anticipated advantage
is in the minimal rate. The minimal rate of \cite{Leinonen2009} is
very close in both versions and is only about \textasciitilde 70\%
of our minimal rate, and decreases with $N$ while our minimal rate
increases. The minimal rate of \cite{Chen2006} converges to zero,
since for $N=K$, the probability that some users will not be allocated
a channel is high. 

To show the tightness of our selected $M$ we now demonstrate the
threshold phenomenon for the existence of a PM. In Table \ref{tab:Existence-of-good},
we present the empirical probability that a PM does not exist for
both $M=\left\lfloor b\ln\left(K\right)\right\rfloor $ and $M=\left\lfloor 1.5\left(b+1\right)\ln\left(K\right)\right\rfloor $,
with $b=4$, averaged over 10000 simulations. For $M=\left\lfloor b\ln\left(K\right)\right\rfloor $,
already for $N=10$, the probability that a PM does not exist was
close to 0.5 and monotonically increased with $N$. This is in agreement
with Lemma \ref{M is not fixed}, that shows that for $M\leq b\ln\left(K\right)$
the probability that a PM does not exist is bounded from below by
0.5, for large enough $N$. For $M=\left\lfloor 1.5\left(b+1\right)\ln\left(K\right)\right\rfloor $,
the empirical probability that a PM does not exist was much below
the bound of Theorem \ref{thm:Main} for all $N$, but the gap between
the two decreased with $N$. This is to be expected, since the proof
of Theorem \ref{thm:Main} uses arguments that hold, and become tight,
for large enough $N$.

\begin{figure}[th]
\begin{centering}
\textsf{\includegraphics[width=7cm,height=5cm]{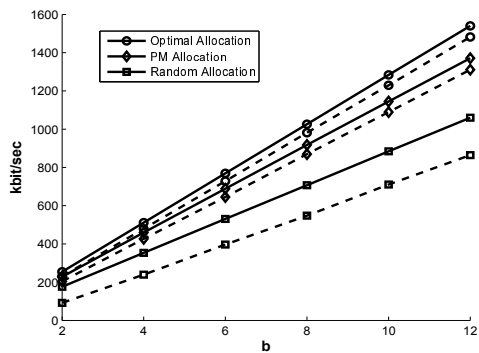}}
\par\end{centering}
\caption{\label{fig:versus b}Achievable rates as a function of $b$ for $N=30$.
Solid lines are used for the mean rates and dashed lines for the minimal
rates.}
\end{figure}

In Figure \ref{fig:versus b}, we present the rates as a function
of $b$ for $N=30$. All the rates increased almost linearly with
$b$, and the ratio of the rates of a PM allocation to the optimal
allocation rates remained almost constant for all $b$. This implies
that the results of Figure \ref{fig:The-sum-rate-of} are similar
for each $b$ value. 

\begin{table*}[t]
\caption{\label{tab:Existence-of-good}Existence of PM allocations as a function
of $N$ and $M$, for $b=4$.}

~~~~~~~~~~~~~~~~%
\begin{tabular}{|c|c|c|c|c|c|}
\hline 
$\Pr\left(\mathcal{E}_{bN}\right)$ & $N=10$ & $N=25$ & $N=50$ & $N=75$ & $N=100$\tabularnewline
\hline 
$M=\left\lfloor b\ln\left(K\right)\right\rfloor $ & 0.44 & 0.52 & 0.54 & 0.65 & 0.67\tabularnewline
\hline 
$M=\left\lfloor 1.5(b+1)\ln\left(K\right)\right\rfloor $ & 0.004 & 0.004 & 0.004 & 0.004 & 0.003\tabularnewline
\hline 
$M=\left\lfloor 1.5(b+1)\ln\left(K\right)\right\rfloor $ upper bound & 0.17 & 0.06 & 0.03 & 0.02 & 0.01\tabularnewline
\hline 
\end{tabular}
\end{table*}

\subsection{Correlated Non-Identically Distributed Channels}

In this subsection, we used correlated resource blocks as the channels,
each consisting of 12 subcarriers. The rates are normalized per subcarrier.
We used the Extended Pedestrian A model (EPA, see \cite{LTE2009})
for the excess tap delay and the relative power of each tap. In this
model, the channel gains of different users are not identically distributed,
which is the practical scenario. We used $N=32,64,96,128$. 

\begin{figure}[th]
\begin{centering}
\textsf{\includegraphics[width=7cm,height=5cm]{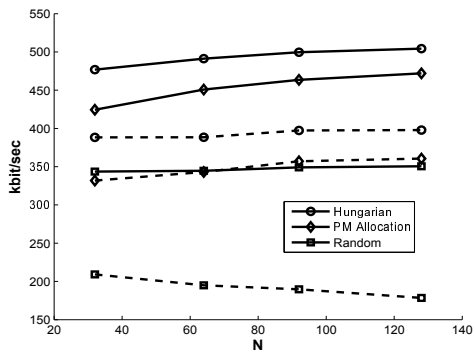}}
\par\end{centering}
\caption{\label{fig:The-sum-rate-of-Correlated}Achievable rates as a function
of $N$ for non-identically distributed correlated channels. Solid
lines represent the mean rates and dashed lines the minimal rates.}
\end{figure}

Figure \ref{fig:The-sum-rate-of-Correlated} shows the mean and minimal
rates as a function of $N$ averaged over 100 realizations for $b=4$.
With correlated channels the diversity decreases; namely, adjacent
channels are likely to be good or bad together. This makes the multiuser
diversity gain lower and as a result the rates are lower, and look
like the rates of the uncorrelated fading for smaller values of $N$.
Additionally, this reduced diversity lead us to choose $M=\left\lceil 2\left(b+1\right)\ln\left(K\right)\right\rceil $,
since a higher $M$ is needed for a PM to exist in the correlated
user-channel graph. Indeed, we always found PMs for this choice of
$M$. However, a larger $M$ makes the PM allocation rates lower.
Nevertheless, these rates are still more than 85\% of optimal rates
already for $N=50$ and get closer as $N$ increases. Given these
experimental results, we conjecture that our analytical results can
be extended to the case of correlated channels.

\begin{figure}[th]
\begin{centering}
\textsf{\includegraphics[width=7cm,height=5cm]{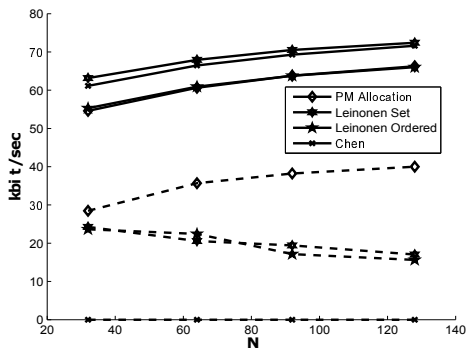}}
\par\end{centering}
\caption{\label{fig:ComparisonCorrelated}Achievable rates as a function of
$N$ for non-identically distributed correlated channels, compared
to state of the art algorithms. Solid lines represent the mean rates
and dashed lines the minimal rates.}
\end{figure}

In Figure \ref{fig:ComparisonCorrelated}, we repeated the above experiment
with $b=1$ and a comparison to \cite{Leinonen2009} and \cite{Chen2006}.
Note that both \cite{Leinonen2009} and \cite{Chen2006} assumed uncorrelated
channels and identically distributed channel gains. In a thresholding
approach, a different threshold can be used for each user, compromising
mean-rate for fairness. However, since the minimal rate of \cite{Chen2006}
converges to zero anyway, we chose the threshold in order to optimize
the mean-rate. As shown, the results are similar to those of Figure
\ref{fig:ComparisonRatesb1}, and the feedback overhead comparison
is of course the same. The main difference is that the minimal rate
of all methods is lower. This does not represent a degradation for
any of the algorithms, since with non-identically distributed channel
gains, the minimal rate user is the user with the worst global conditions.
In our method, the minimal rate is still significantly better than
those of \cite{Leinonen2009} and \cite{Chen2006}, since this minimal
rate user got one of his best channels (even though all of them were
relatively bad).

\subsection{Unequal Channel Allocation}

\begin{figure}[th]
\begin{centering}
\textsf{\includegraphics[width=6cm,height=6cm]{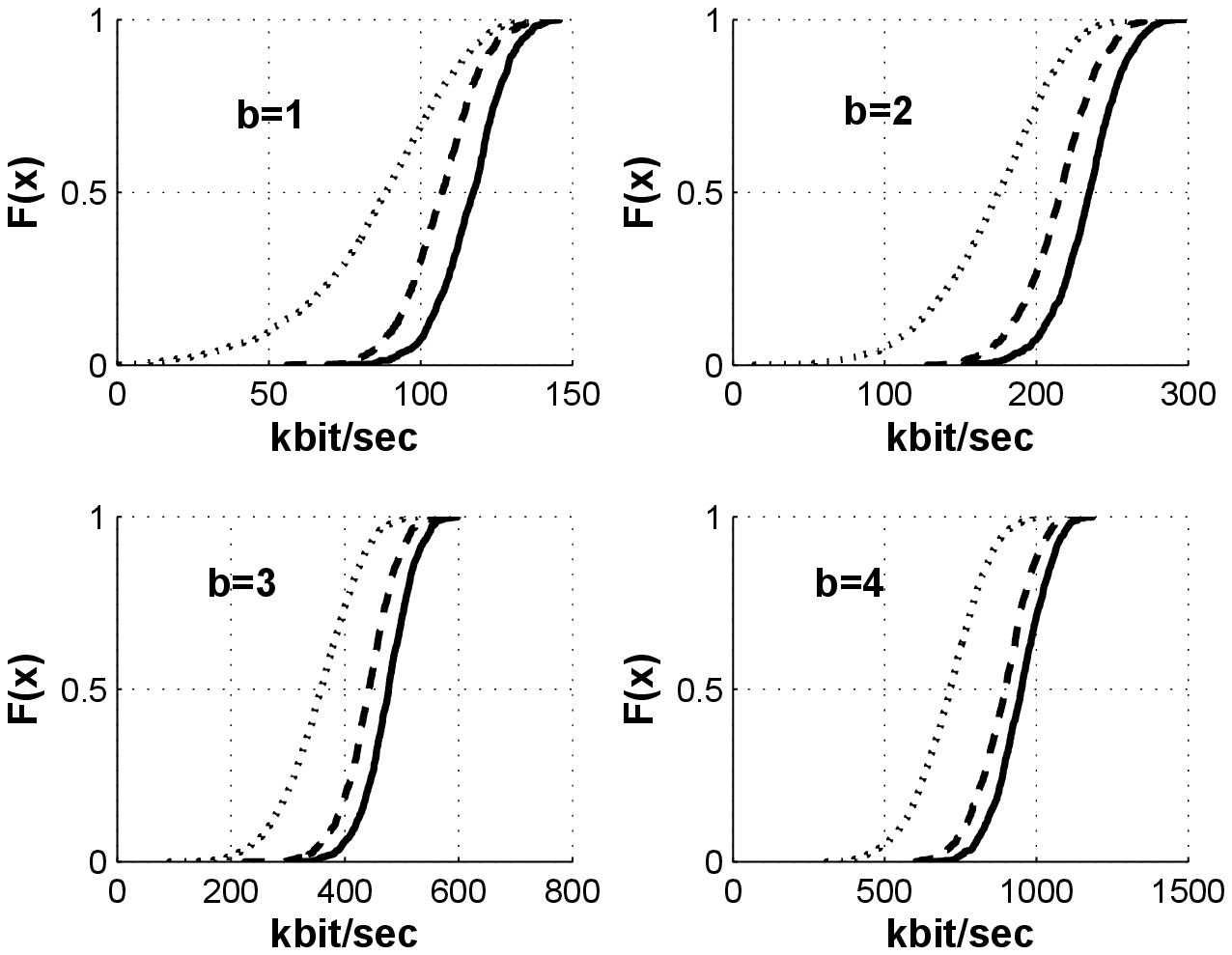}\includegraphics[width=6cm,height=6cm]{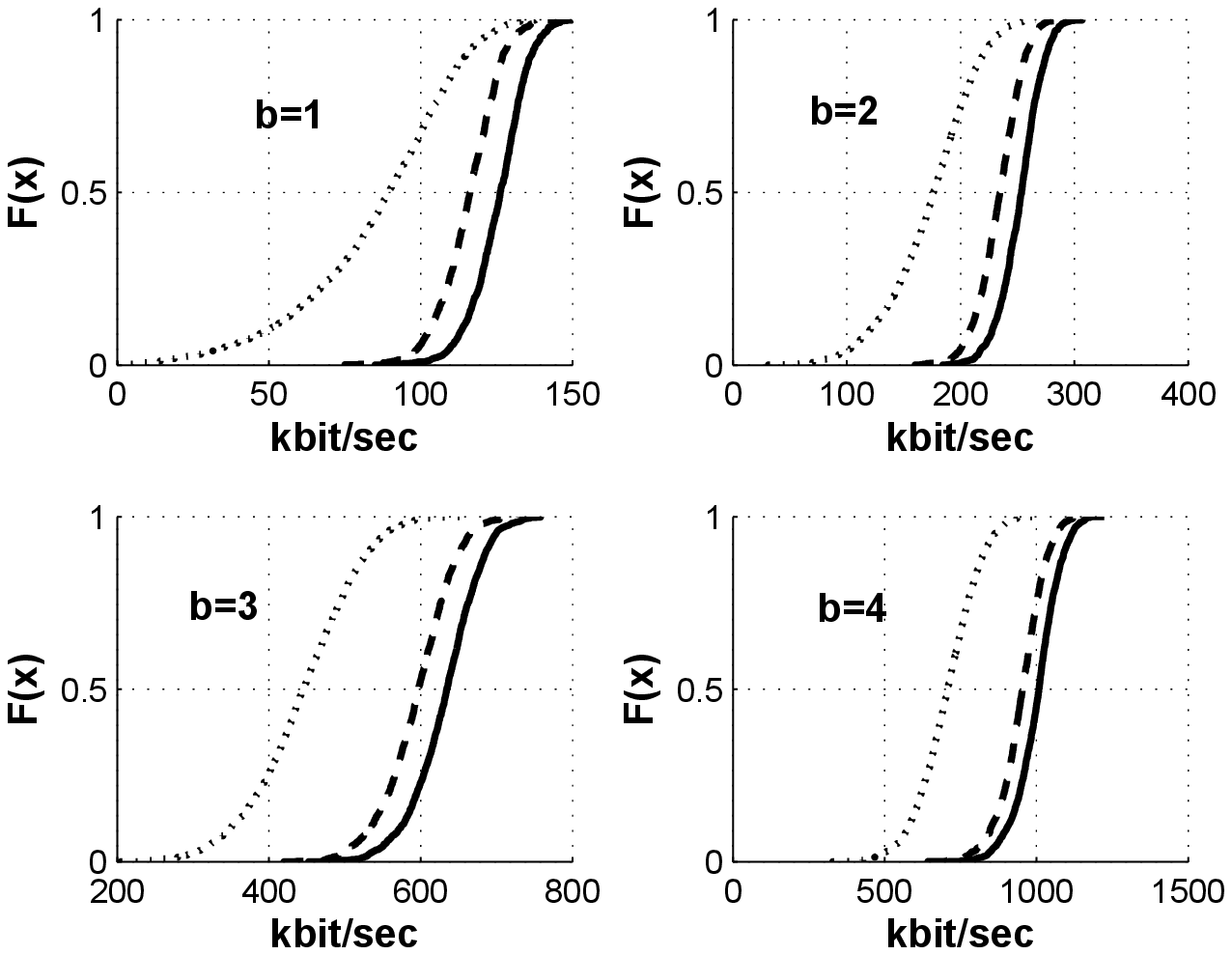}}
\par\end{centering}
\caption{\label{fig:The-sum-rate-different b}Empirical CDFs of the rates for
different classes of users with $N=100$, for uncorrelated channels
(left) and with $N=128$, for correlated resource blocks (right).
Solid lines are used for optimal rates, dashed lines for the PM allocation
rates and dotted lines for the random allocation rates. }
\end{figure}

In this subsection, we allocated users a different number of channels,
by dividing them into four classes. Specifically, $\frac{N}{4}$ of
the users were allocated a single channel, $\frac{N}{4}$ two channels,
$\frac{N}{4}$ three channels and $\frac{N}{4}$ four channels. We
used $M=\left\lceil 3.75\ln\left(K\right)\right\rceil $ for all users,
which amounts to using the average $b$ in $M=\left\lceil 1.07\left(b+1\right)\ln\left(K\right)\right\rceil $.
Such a differentiation might occur if different users require a different
quality of service (QoS). We simulated 100 realizations for each value
from $N=12,20,48,72,100$. PMs still existed in all realizations.
We ran another 100 realizations with correlated resource blocks, in
the same manner as in Figure \ref{fig:The-sum-rate-of-Correlated}
and used $M=\left\lceil 5\ln\left(K\right)\right\rceil $ for all
users. Figure \ref{fig:The-sum-rate-different b} shows the empirical
CDFs of the rates for the four classes, for uncorrelated channels
and correlated resource blocks, with $N=100$ and $N=128$. Note that
the PM allocation rates are quite concentrated and most of the users
from the same class get similar rates. These rates are always better
than those of a random allocation, and close to the optimal allocation
rates. These results confirm our belief that our analysis can be generalized
to the case of an unequal allocation, which is further supported by
our proof itself. 

\subsection{Water-Filling Gain}

\begin{figure}[th]
\begin{centering}
\textsf{\includegraphics[width=7cm,height=5cm]{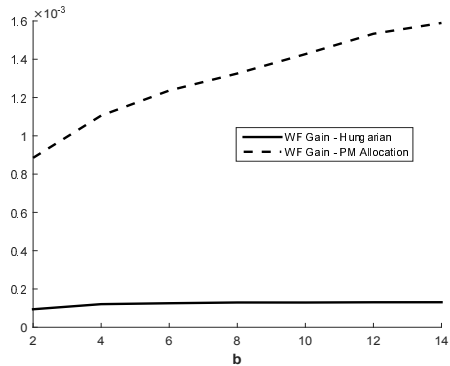}}
\par\end{centering}
\caption{\label{fig:WFGain}The water-filling relative mean-rate gain $\frac{R_{WF}-R}{R}$
as a function of $b$ for $N=30$ and an SNR of 10dB.}
\end{figure}

In all the above results, water-filling over the $b$ channels of
each user had a negligible gain. To be exact, it never improved any
of the rates by more than 0.01 kbps. This is consistent with our analytical
result in Corollary \ref{cor:WF}. We repeated our experiment for
an SNR of 10dB, $N=30$ and $M=\left\lceil 2\left(b+1\right)\ln\left(K\right)\right\rceil $,
where the water-filling gain should be more significant than with
our original parameters. We used $b=2,4,6,8,10,12,14$. In Figure
\ref{fig:WFGain} we present the water filling relative mean-rate
gain, defined as $\frac{R_{WF}-R}{R}$, as a function of $b$. Clearly,
even with an SNR of 10dB, water-filling did not improve the mean rates
by more than 0.16\% for either the Hungarian algorithm or our PM allocation
for any $b$. 

\section{Conclusions}

In this paper we suggested an approach aimed at achieving the multiuser
diversity gain for resource block allocation under the frequency-selective
channel. Our approach requires knowledge of the indices of the $M$-best
channels for each of the $N$ users, instead of the whole $NK$ channel
gains. 

We showed that $M=\left(b+1\right)\left(1+\varepsilon\right)\ln\left(K\right)$
for some $\varepsilon>0$ guarantees that a PM allocation, where each
user gets all his $b$ channels from his $M$-best channels, asymptotically
exists. For this value of $M$, our suboptimal approach was shown
to be asymptotically optimal for a broad class of fading distributions,
both in the sum-rate and max-min senses.

Our algorithm is the first limited feedback algorithm that has provably
asymptotic optimality both of the sum-rate and the minimal rate. It
also requires significantly less feedback than the state of the art
algorithms that achieve a good sum-rate. Hence, it constitutes a major
step toward achieving the multiuser diversity gain of a selective
channel in practice.

A future direction could involve trying to lift the assumption of
i.i.d. channel gains by only assuming $m$-dependent channel gains
and using the techniques presented in \cite{Bistritz2017}. 

\appendices{}

\section{Proofs\label{sec:Proofs}}

For our results, we need the following two lemmas.
\begin{lem}
\label{prop:simple}For each $0\leq l\leq K$ and $0\leq M\leq l$
we have $\frac{\binom{l}{M}}{\binom{K}{M}}\leq\left(\frac{l}{K}\right)^{M}$.
\end{lem}
\begin{IEEEproof}
\begin{equation}
\binom{l}{M}\binom{K}{M}^{-1}=\frac{\frac{l!}{\left(l-M\right)!}}{\frac{K!}{\left(K-M\right)!}}=\prod\limits _{i=1}^{M}\frac{l-M+i}{K-M+i}=\prod\limits _{j=0}^{M-1}\frac{l-j}{K-j}\underset{(a)}{\leq}\left(\frac{l}{K}\right)^{M}\label{eq:6-1}
\end{equation}
where (a) follows since if $l<K$ then $f\left(x\right)=\frac{l-x}{K-x}$
is monotonically decreasing for all $x<K$, so $\frac{l-j}{K-j}\leq\frac{l}{K}$
for each $0\leq j<M$. 
\end{IEEEproof}
\begin{lem}
\label{lem:Ugly}Let $M=\left(b+1\right)\left(1+\varepsilon\right)\ln K$
for some $\varepsilon>0$. Define the function 
\begin{equation}
f\left(x\right)=K\left(1+\frac{1}{b}\right)h_{b}\left(\frac{x+1}{K}\right)\ln2-\frac{M}{Kb}\left(K-x\right)\left(x+1\right)\label{eq:11-1}
\end{equation}
where $h_{b}$ is the binary entropy function (see \cite[page 666]{Cover2012}).
Then, for large enough $K$, we have for all $M\leq x\leq K-1$
\begin{equation}
f\left(x\right)\leq3-\left(1+\frac{1}{b}\right)\left(1+\frac{3}{2}\varepsilon\right)\ln K\label{eq:11-2}
\end{equation}
\end{lem}
\begin{IEEEproof}
Since $\frac{dKh_{b}\left(\frac{x+1}{K}\right)}{dx}=\log_{2}\left(\frac{K}{x+1}-1\right)$
we have
\begin{equation}
\frac{df\left(x\right)}{dx}=\left(1+\frac{1}{b}\right)\ln\left(\frac{K}{x+1}-1\right)-\frac{M}{b}\left(1-\frac{1}{K}\right)+\frac{2}{b}x\frac{M}{K}\label{eq:11}
\end{equation}
 and
\begin{equation}
\frac{d^{2}f\left(x\right)}{dx^{2}}=-\frac{1+\frac{1}{b}}{K-x-1}-\frac{1+\frac{1}{b}}{x+1}+\frac{2}{b}\frac{M}{K}.\label{eq:12}
\end{equation}
so for any $x\in\left[\frac{\left(b+1\right)K}{M}-1,K-\frac{\left(b+1\right)K}{M}-1\right]$
we have $\frac{d^{2}f\left(x\right)}{dx^{2}}\geq0$. For $K-\frac{\left(b+1\right)K}{M}-1\leq x\leq K-2$
we have 
\begin{equation}
\frac{df\left(x\right)}{dx}\geq\left(1-\frac{2b+2}{M}-\frac{1}{K}\right)\frac{M}{b}-\left(1+\frac{1}{b}\right)\ln\left(K-1\right)\underset{\left(a\right)}{\geq}0\label{eq:13}
\end{equation}
where (a) follows for $M=\left(b+1\right)\left(1+\varepsilon\right)\ln\left(K\right)$
and a large enough $K$. For $M\leq x\leq\frac{\left(b+1\right)K}{M}-1$
we have
\begin{equation}
\frac{df\left(x\right)}{dx}\leq\left(1+\frac{1}{b}\right)\ln\left(\frac{K}{M}-1\right)-\frac{M}{b}\left(1-\frac{1}{K}\right)+2+\frac{2}{b}\underset{\left(a\right)}{\leq}0.\label{eq:14}
\end{equation}
where (a) holds for $M=\left(b+1\right)\left(1+\varepsilon\right)\ln\left(K\right)$
and a large enough $K$. Thus, $f\left(x\right)$ is convex in $\left[\frac{\left(b+1\right)K}{M}-1,K-\frac{\left(b+1\right)K}{M}-1\right]$,
increasing in $\left[K-\frac{\left(b+1\right)K}{M}-1,K-2\right]$
and decreasing in $\left[M,\frac{\left(b+1\right)K}{M}-1\right]$.
Hence, the maximum of $f\left(x\right)$ must be attained at the borders
$l=M,K-2$. Observe that for all $l<\frac{K}{2}$
\begin{equation}
Kh_{b}\left(\frac{l}{K}\right)=l\log_{2}\frac{K}{l}-\left(K-l\right)\log_{2}\left(1-\frac{l}{K}\right)\underset{(a)}{\leq}2l+l\log_{2}\frac{K}{l}\label{eq:15}
\end{equation}
 where in (a) we used $\log_{2}\left(1-\frac{l}{K}\right)\geq-\frac{2l}{K}$
which is true for $l\leq\frac{K}{2}$. Now we substitute $x=M,K-2$
in $f\left(x\right)$ to obtain

\begin{multline}
f\left(M\right)\underset{(a)}{\leq}\left(1+\frac{1}{b}\right)\left(2\left(M+1\right)\ln2+\left(M+1\right)\ln\frac{K}{M+1}\right)-\frac{M^{2}}{b}\left(1-\frac{1}{K}\right)+\frac{M^{3}}{bK}-\frac{M}{b}\underset{(b)}{\leq}\\
M\left(1+\frac{1}{b}\right)\left(3+\left(1+\frac{1}{M}\right)\ln K-\frac{M}{b+1}\left(1-\frac{1}{K}\right)\right)\label{eq:16}
\end{multline}
where (a) is from \eqref{eq:15} and (b) is for large enough $K$
due to $M=\left(b+1\right)\left(1+\varepsilon\right)\ln\left(K\right)$.
Next
\begin{multline}
f\left(K-2\right)\underset{\left(a\right)}{=}\ln\left(2\right)\left(1+\frac{1}{b}\right)Kh_{b}\left(\frac{1}{K}\right)-\frac{M}{b}\frac{\left(K-1\right)\left(K-2\right)}{K}+\frac{M}{b}\frac{\left(K-2\right)^{2}}{K}-\frac{M}{b}\underset{(b)}{\leq}\\
\left(1+\frac{1}{b}\right)\ln4K-\frac{2M}{b}\left(1-\frac{1}{K}\right)\label{eq:17}
\end{multline}
where (a) is from $h_{b}\left(1-\frac{1}{K}\right)=h_{b}\left(\frac{1}{K}\right)$
and (b) from \eqref{eq:15}. Hence for large enough $K$ \eqref{eq:17}
is larger than \eqref{eq:16}. We conclude that for all $M\leq x\leq K-1$
\begin{multline}
f\left(x\right)\leq\underset{x}{\max}f\left(x\right)\leq\left(1+\frac{1}{b}\right)\ln4K-\frac{2M}{b}\left(1-\frac{1}{K}\right)\\
\leq3-\left(1+\frac{1}{b}\right)\ln K\left(2\left(1+\varepsilon\right)\left(1-\frac{1}{K}\right)-1\right)\underset{(a)}{\leq}3-\left(1+\frac{1}{b}\right)\left(1+\frac{3}{2}\varepsilon\right)\ln K\label{eq:18}
\end{multline}
where (a) is true for large enough $K$ such that $\frac{1}{K}\leq\frac{1}{4}\frac{\varepsilon}{1+\varepsilon}$. 
\end{IEEEproof}

\subsection{Proof of Theorem \ref{thm:Main}}
\begin{IEEEproof}
The bipartite graph can also be represented using a $K\times K$ preferences
matrix $U$. Our matrix $U$ is a simplified binary matrix where $u_{n,k}=0$
if channel $k$ is one of the $M$-best channels of agent $n$ and
$u_{n,k}=1$ otherwise. It consists of $\frac{K}{b}$ submatrices
of $b$ identical rows each, and exactly $M$ zeros at each row. We
want to show that the probability that the Hungarian algorithm on
$U$ converges after one iteration approaches one as $K\rightarrow\infty$;
hence the resulting cost is zero, which represents a PM in the equivalent
bipartite graph. Note that since $K=bN$ for a positive integer $b$,
when $K\rightarrow\infty$ then also $N\rightarrow\infty$. Denote
by $Q_{l}\left(K,M\right)$ the probability that $U$ can be covered
by $K-l-1$ rows and $l$ columns such that all its zeros are covered,
and that there is no smaller $l$ with this property. We want to upper
bound $Q_{l}\left(K,M\right)$ for each $l$. Now observe that if
$U$ is covered such that a row is covered but one of its identical
rows is not, this means that $U$ could be covered without this row
cover. Hence, if the row covers do not include all the identical rows,
they are redundant. This means that only values of $K-l-1$ (or $l+1$)
that are a multiple of $b$ should be considered, and $Q_{l}\left(K,M\right)=0$
for all other values. We proceed by counting the number of combinations
of each such matrix
\begin{equation}
Q_{l}\left(K,M\right)\leq\underbrace{\binom{K}{l}\binom{\frac{K}{b}}{\frac{l+1}{b}}}_{A}\underbrace{\binom{K}{M}^{\frac{K-l-1}{b}}}_{B}\underbrace{\binom{l}{M}^{\frac{l+1}{b}}}_{C}\binom{K}{M}^{-\frac{K}{b}}=\binom{K}{l}\binom{\frac{K}{b}}{\frac{l+1}{b}}\left(\frac{\binom{l}{M}}{\binom{K}{M}}\right)^{\frac{l+1}{b}}\label{eq:5}
\end{equation}
where $A$ is the number of combinations for the index of the $K-l-1$
covered rows and the $l$ covered columns. Now we have to count the
combinations for the structure of the covered and uncovered rows.
The factor $B$ is the number of combinations for $K-l-1$ covered
rows with exactly $M$ zeros in each. We are left with $l+1$ rows
that we know can be covered by $l$ columns, and no less than $l$.
Hence each of these rows has zeros only in these $l$ columns, so
there can be no more than $C$ such rows. Finally, the denominator
$\binom{K}{M}^{\frac{K}{b}}$ is the number of combinations of a matrix
with $M$ zeros in each row and $b$ duplicates of each such row,
that constitute the whole probability space. 

For $l=K-1$ we get
\begin{equation}
Q_{K-1}\left(K,M\right)\underset{(a)}{\leq}K\left(1-\frac{1}{K}\right)^{\frac{M}{b}K}\underset{(b)}{\leq}Ke^{-\frac{M}{b}}\label{eq:8}
\end{equation}
where (a) follows from Lemma \ref{prop:simple} and (b) from the inequality
$\left(1-\frac{1}{K}\right)^{K}\leq e^{-1}$. Note that for $l=1,...,M-1$
it is trivial that $Q_{l}\left(K,M\right)=0$ since each row has exactly
$M$ zeros. For the rest of the proof we assume that $M<l\leq K-2$.
For these values, we have
\begin{equation}
\binom{K}{l}\binom{\frac{K}{b}}{\frac{l+1}{b}}=\frac{l+1}{K-l}\binom{K}{l+1}\binom{\frac{K}{b}}{\frac{l+1}{b}}\underset{(a)}{\leq}\frac{\sqrt{b}}{2\pi}\frac{1}{\left(K-l\right)\left(1-\frac{l+1}{K}\right)}2^{K\left(1+\frac{1}{b}\right)h_{b}\left(\frac{l+1}{K}\right)}\label{eq:7}
\end{equation}
where in (a) we used $\binom{K}{l}\leq\sqrt{\frac{1}{2\pi l\left(1-\frac{l}{K}\right)}}2^{Kh_{b}\left(\frac{l}{K}\right)}$,
which is valid for all $0<l<K$ (where $h_{b}$ is the binary entropy
function, see \cite[page 666]{Cover2012}). Combining the above we
obtain from \eqref{eq:5} the following upper bound
\begin{equation}
Q_{l}\left(K,M\right)\underset{(a)}{\leq}\frac{\sqrt{b}2^{K\left(1+\frac{1}{b}\right)h_{b}\left(\frac{l+1}{K}\right)}}{2\pi\left(K-l\right)\left(1-\frac{l+1}{K}\right)}\left(\frac{l}{K}\right)^{\frac{M}{b}\left(l+1\right)}\underset{(b)}{\leq}\frac{\sqrt{b}}{2\pi\left(K-l\right)\left(1-\frac{l+1}{K}\right)}e^{\ln\left(2\right)K\left(1+\frac{1}{b}\right)h_{b}\left(\frac{l+1}{K}\right)-\frac{M}{Kb}\left(K-l\right)\left(l+1\right)}\label{eq:9}
\end{equation}
where (a) is due to Lemma \ref{prop:simple} and \eqref{eq:7} and
(b) is due to $1-\frac{K-l}{K}\leq e^{-\frac{K-l}{K}}$. We want to
find the maximal value of this bound as a function of $l$. Define
the auxiliary function $f\left(x\right)$ as in \eqref{eq:11-1} and
denote by $\mathcal{E}_{K}$ the event in which a PM does not exist.
We conclude by the union bound that
\begin{multline}
\Pr\left(\mathcal{E}_{K}\right)\leq Q_{K-1}\left(K,M\right)+\sum_{l=M}^{K-2}Q_{l}\left(K,M\right)\leq Ke^{-\frac{M}{b}}+\sum_{l=M}^{K-2}\frac{\sqrt{b}K}{2\pi\left(K-l\right)\left(K-l-1\right)}e^{f\left(l\right)}\underset{\left(a\right)}{\leq}\\
Ke^{-\frac{M}{b}}+\frac{\frac{\sqrt{b}}{2\pi}e^{3}}{K^{\frac{3}{2}\varepsilon+\frac{\frac{3}{2}\varepsilon+1}{b}}}\sum_{l=M}^{K-2}\frac{1}{\left(K-l-1\right)^{2}}\underset{(b)}{\leq}Ke^{-\frac{M}{b}}+\frac{\frac{\sqrt{b}}{2\pi}e^{3}}{K^{\frac{3}{2}\varepsilon+\frac{\frac{3}{2}\varepsilon+1}{b}}}\sum_{k=1}^{\infty}\frac{1}{k^{2}}\underset{(c)}{\leq}\frac{1}{K^{\varepsilon+\frac{\varepsilon+1}{b}}}+\frac{\pi\sqrt{b}e^{3}}{12}\frac{1}{K^{\frac{3}{2}\varepsilon+\frac{\frac{3}{2}\varepsilon+1}{b}}}\label{eq:19}
\end{multline}
where (a) follows from Lemma \ref{lem:Ugly} for large enough $K$,
(b) follows by substituting $k=K-l-1$ and adding terms to the sum.
Inequality (c) follows from \eqref{eq:8} and $\sum_{k=1}^{\infty}\frac{1}{k^{2}}=\frac{\pi^{2}}{6}$.
Note that if $\varepsilon\geq1$ then $\sum_{K=1}^{\infty}\Pr\left(\mathcal{E}_{K}\right)<\infty$.
Hence, by the Borel-Cantelli lemma, the probability that an infinite
number of events from $\left\{ \mathcal{E}_{K}\right\} $ will occur
is zero. Hence, the indicator of $\mathcal{E}_{K}$ converges almost
surely to zero. 
\end{IEEEproof}

\subsection{Proof of Lemma \ref{M is not fixed}}
\begin{IEEEproof}
Let $\mathcal{E}_{0,k}$ be the event in which channel $k$ is not
one of the $M$-best channels for any of the users. We bound from
below and above the probability that there exists a channel that is
not good for any of the users; i.e., the probability of $\mathcal{E}_{0}=\bigcup_{k}\mathcal{E}_{0,k}$.
Due to the i.i.d. assumption on the channel gains and their independency
between users we have $\Pr\left(\mathcal{E}_{0,k}\right)=\left(1-\frac{M}{K}\right)^{N}$.
From the union bound we obtain that
\begin{equation}
\Pr\left(\mathcal{E}_{0}\right)=\Pr\left(\bigcup_{k}\mathcal{E}_{0,k}\right)\leq K\left(1-\frac{M}{K}\right)^{N}.\label{eq:3-1}
\end{equation}
Since $\left(1-\frac{M}{K}\right)^{N}$ decreases with $M$, using
$M\geq b\left(1+\varepsilon\right)\ln K$ leads to
\begin{equation}
\underset{K\rightarrow\infty}{\lim}\Pr\left(\mathcal{E}_{0}\right)\leq\underset{K\rightarrow\infty}{\lim}K\left(1-\frac{M}{K}\right)^{N}\leq\underset{K\rightarrow\infty}{\lim}K\left(1-\frac{b\left(1+\varepsilon\right)\ln K}{K}\right)^{\frac{K}{b}}=\underset{K\rightarrow\infty}{\lim}Ke^{-\left(1+\varepsilon\right)\ln K}=0.\label{eq:4}
\end{equation}
Now we want to bound from above the probability of a PM. We do so
by using the inclusion-exclusion principle. We use the following lower
bound
\begin{equation}
\Pr\left(\bigcup_{k}\mathcal{E}_{0,k}\right)\geq\sum_{k}\Pr\left(\mathcal{E}_{0,k}\right)-\sum_{k_{1},k_{2}}\Pr\left(\mathcal{E}_{0,k_{1}}\bigcap\mathcal{E}_{0,k_{2}}\right).\label{eq:57-2}
\end{equation}
By direct counting of the number of user-channel graphs with two undesired
channels we obtain
\begin{equation}
\sum_{k_{1},k_{2}}\Pr\left(\mathcal{E}_{0,k_{1}}\bigcap\mathcal{E}_{0,k_{2}}\right)=\left(\begin{array}{c}
K\\
2
\end{array}\right)\left(\begin{array}{c}
K-2\\
M
\end{array}\right)^{N}\left(\begin{array}{c}
K\\
M
\end{array}\right)^{-N}\underset{(a)}{\leq}\frac{K^{2}}{2}\left(1-\frac{2}{K}\right)^{M\frac{K}{b}}\label{eq:58}
\end{equation}
since $\binom{K}{2}$ is the number of choices for these two channels,
$\binom{K-2}{M}$ is the number of combinations of the edges of each
agent and $\binom{K}{M}^{N}$ is the total number of user-channel
graphs. Inequality (a) is from Lemma \ref{prop:simple}. We obtain
\begin{equation}
\Pr\left(\mathcal{E}_{0}\right)=\Pr\left(\bigcup_{k}\mathcal{E}_{0,k}\right)\geq K\left(1-\frac{M}{K}\right)^{N}-\frac{K^{2}}{2}\left(1-\frac{2}{K}\right)^{M\frac{K}{b}}.\label{eq:59}
\end{equation}
Obviously, $\Pr\left(\mathcal{E}_{0}\right)$ decreases with $M$.
Hence, using $M=b\ln K$, we conclude that
\begin{equation}
\underset{K\rightarrow\infty}{\lim}\Pr\left(\mathcal{E}_{0}\right)\geq\underset{K\rightarrow\infty}{\lim}\left(K\underbrace{\left(1-\frac{b\ln K}{K}\right)^{\frac{K}{b}}}_{\rightarrow e^{-\ln K}}-\frac{K^{2}}{2}\underbrace{\left(1-\frac{2}{K}\right)^{K\ln K}}_{\rightarrow e^{-2\ln K}}\right)=\frac{1}{2}.\label{eq:60}
\end{equation}
\end{IEEEproof}

\subsection{Proof of Proposition \ref{prop:The-Rayleigh,--Nakagami,}}
\begin{IEEEproof}
For the Rayleigh distribution we have $F_{X}\left(x\right)=1-e^{-\frac{x^{2}}{2\sigma^{2}}}$,
so we simply have $\frac{1-F_{X}\left(x\right)}{\alpha x^{\beta}e^{-\lambda x^{\gamma}}}=1$
for $\alpha=1$, $\beta=0$,$\gamma=2$ and $\lambda=\frac{1}{2\sigma^{2}}$.
Denote the PDF by $f_{X}\left(x\right)$. From l'Hôpital's \cite[Page 109]{Rudin1964}
rule we obtain
\begin{equation}
\underset{x\rightarrow\infty}{\lim}\frac{1-F_{X}\left(x\right)}{\alpha x^{\beta}e^{-\lambda x^{\gamma}}}=\underset{x\rightarrow\infty}{\lim}\frac{f_{X}\left(x\right)}{\left(\alpha\lambda\gamma-\frac{\alpha\beta}{x^{\gamma}}\right)x^{\beta-1+\gamma}e^{-\lambda x^{\gamma}}}.\label{eq:21-1}
\end{equation}
For an $m$-Nakagami distribution with parameter $\Omega>0$, $f_{X}\left(x\right)=\frac{2m^{m}}{\Gamma\left(m\right)\Omega^{m}}x^{2m-1}e^{-\frac{m}{\Omega}x^{2}}$.
Substituting $\alpha=\frac{m^{m-1}}{\Gamma\left(m\right)\Omega^{m-1}}$,
$\beta=2m-2$,$\gamma=2$ and $\lambda=\frac{m}{\Omega}$ in \eqref{eq:21-1}
yields
\begin{equation}
\underset{x\rightarrow\infty}{\lim}\frac{\frac{2m^{m}}{\Gamma\left(m\right)\Omega^{m}}x^{2m-1}e^{-\frac{m}{\Omega}x^{2}}}{\left(\frac{2m^{m}}{\Gamma\left(m\right)\Omega^{m}}-\frac{\alpha\beta}{x^{2}}\right)x^{2m-1}e^{-\frac{m}{\Omega}x^{2}}}=1.\label{eq:20-3}
\end{equation}
For the Rice distribution (with an expectation shifted by a constant
$v$) $f_{X}\left(x\right)=f_{Y}\left(x+v\right)=\frac{x+v}{\sigma^{2}}e^{-\frac{\left(x+v\right)^{2}+v^{2}}{2\sigma^{2}}}I_{0}\left(\frac{\left(x+v\right)v}{\sigma^{2}}\right)$.
Substituting $\alpha=\sigma\sqrt{\frac{1}{2\pi v}}$, $\beta=-\frac{1}{2}$,$\gamma=2$
and $\lambda=\frac{1}{2\sigma^{2}}$ in \eqref{eq:21-1} yields
\begin{equation}
\underset{x\rightarrow\infty}{\lim}\frac{\frac{x+v}{\sigma^{2}}e^{-\frac{\left(x+v\right)^{2}+v^{2}}{2\sigma^{2}}}I_{0}\left(\frac{\left(x+v\right)v}{\sigma^{2}}\right)}{\left(\frac{1}{\sigma}\sqrt{\frac{1}{2\pi v}}+\frac{\alpha}{2x^{2}}\right)\sqrt{x}e^{-\frac{x^{2}}{2\sigma^{2}}}}\underset{(a)}{=}\underset{x\rightarrow\infty}{\lim}\frac{\sqrt{x+v}}{\sqrt{x}}=1\label{eq:20-4}
\end{equation}
where (a) follows since $\underset{x\rightarrow\infty}{\lim}\frac{I_{0}\left(\frac{xv}{\sigma^{2}}\right)}{\frac{\sigma}{\sqrt{2\pi xv}}e^{\frac{xv}{\sigma^{2}}}}=1$.
For a Gaussian distribution $f_{X}\left(x\right)=\frac{1}{\sqrt{2\pi\sigma^{2}}}e^{-\frac{1}{2\sigma^{2}}x^{2}}$.
Substituting $\alpha=\sqrt{\frac{\sigma^{2}}{2\pi}}$, $\beta=-1$,$\gamma=2$
and $\lambda=\frac{1}{2\sigma^{2}}$ in \eqref{eq:21-1} yields
\begin{equation}
\underset{x\rightarrow\infty}{\lim}\frac{\frac{1}{\sqrt{2\pi\sigma^{2}}}e^{-\frac{1}{2\sigma^{2}}x^{2}}}{\left(\frac{1}{\sqrt{2\pi\sigma^{2}}}+\frac{\alpha}{x^{2}}\right)e^{-\frac{1}{2\sigma^{2}}x^{2}}}=1.\label{eq:20-5}
\end{equation}
\end{IEEEproof}
\newpage{}

\bibliographystyle{IEEEtran}
\bibliography{twcreferences}

\begin{thebibliography}{10}
\providecommand{\url}[1]{#1}
\csname url@samestyle\endcsname
\providecommand{\newblock}{\relax}
\providecommand{\bibinfo}[2]{#2}
\providecommand{\BIBentrySTDinterwordspacing}{\spaceskip=0pt\relax}
\providecommand{\BIBentryALTinterwordstretchfactor}{4}
\providecommand{\BIBentryALTinterwordspacing}{\spaceskip=\fontdimen2\font plus
\BIBentryALTinterwordstretchfactor\fontdimen3\font minus
  \fontdimen4\font\relax}
\providecommand{\BIBforeignlanguage}[2]{{%
\expandafter\ifx\csname l@#1\endcsname\relax
\typeout{** WARNING: IEEEtran.bst: No hyphenation pattern has been}%
\typeout{** loaded for the language `#1'. Using the pattern for}%
\typeout{** the default language instead.}%
\else
\language=\csname l@#1\endcsname
\fi
#2}}
\providecommand{\BIBdecl}{\relax}
\BIBdecl

\bibitem{bistritz2018efficient}
I.~Bistritz and A.~Leshem, ``Efficient and asymptotically optimal resource
  block allocation,'' in \emph{2018 IEEE Wireless Communications and Networking
  Conference (WCNC)}.\hskip 1em plus 0.5em minus 0.4em\relax IEEE, 2018, pp.
  1--6.

\bibitem{Zhao2007}
Q.~Zhao and B.~M. Sadler, ``A survey of dynamic spectrum access,'' \emph{Signal
  Processing Magazine, IEEE}, vol.~24, no.~3, pp. 79--89, 2007.

\bibitem{Hasan2013}
M.~Hasan, E.~Hossain, and D.~Niyato, ``Random access for machine-to-machine
  communication in {LTE}-advanced networks: issues and approaches,'' \emph{IEEE
  Communications Magazine}, vol.~51, no.~6, pp. 86--93, 2013.

\bibitem{Akkarajitsakul2011}
K.~Akkarajitsakul, E.~Hossain, D.~Niyato, and D.~I. Kim, ``Game theoretic
  approaches for multiple access in wireless networks: A survey,'' \emph{IEEE
  Communications Surveys \& Tutorials}, vol.~13, no.~3, pp. 372--395, 2011.

\bibitem{Seong2006}
K.~Seong, M.~Mohseni, and J.~M. Cioffi, ``Optimal resource allocation for
  {O}{F}{D}{M}{A} downlink systems,'' in \emph{2006 IEEE International
  Symposium on Information Theory}, 2006.

\bibitem{Zhao2015}
W.~Zhao, S.~Wang, and J.~Guo, ``Efficient resource allocation for {OFDMA}-based
  device-to-device communication underlaying cellular networks,'' in
  \emph{Communications in China (ICCC), 2015 IEEE/CIC International Conference
  on}, 2015.

\bibitem{Ghosh2010a}
A.~Ghosh, J.~Zhang, J.~G. Andrews, and R.~Muhamed, \emph{Fundamentals of
  {L}{T}{E}}.\hskip 1em plus 0.5em minus 0.4em\relax Pearson Education, 2010.

\bibitem{Yoo2005}
T.~Yoo and A.~Goldsmith, ``Optimality of zero-forcing beamforming with
  multiuser diversity,'' in \emph{Communications, 2005. ICC 2005. 2005 IEEE
  International Conference on}, 2005.

\bibitem{Guey2004}
J.-C. Guey and L.~D. Larsson, ``Modeling and evaluation of {MIMO} systems
  exploiting channel reciprocity in {TDD} mode,'' in \emph{Vehicular Technology
  Conference, 2004. VTC2004-Fall. 2004 IEEE 60th}, vol.~6, 2004, pp.
  4265--4269.

\bibitem{Gesbert2004}
D.~Gesbert and M.-S. Alouini, ``How much feedback is multi-user diversity
  really worth?'' in \emph{Communications, 2004 IEEE International Conference
  on}, vol.~1, 2004, pp. 234--238.

\bibitem{Sanayei2007}
S.~Sanayei and A.~Nosratinia, ``Opportunistic downlink transmission with
  limited feedback,'' \emph{IEEE Transactions on Information Theory}, vol.~53,
  no.~11, pp. 4363--4372, 2007.

\bibitem{Chen2006}
J.~Chen, R.~A. Berry, and M.~L. Honig, ``Large system performance of downlink
  {OFDMA} with limited feedback,'' in \emph{Information Theory, 2006 IEEE
  International Symposium on}, 2006, pp. 1399--1403.

\bibitem{Chen2008}
------, ``Limited feedback schemes for downlink {OFDMA} based on sub-channel
  groups,'' \emph{IEEE Journal on Selected Areas in Communications}, vol.~26,
  no.~8, pp. 1451--1461, 2008.

\bibitem{Leinonen2009}
J.~Leinonen, J.~H{\"a}m{\"a}l{\"a}inen, and M.~Juntti, ``Performance analysis
  of downlink {OFDMA} resource allocation with limited feedback,'' \emph{IEEE
  Transactions on Wireless Communications}, vol.~8, no.~6, pp. 2927--2937,
  2009.

\bibitem{Erdos1964}
P.~Erdos and A.~Renyi, ``On random matrices,'' \emph{Magyar Tud. Akad. Mat.
  Kutat{\'o} Int. K{\"o}zl}, vol.~8, no. 455-461, p. 1964, 1964.

\bibitem{Yaacoub2012}
E.~Yaacoub and Z.~Dawy, ``A survey on uplink resource allocation in {OFDMA}
  wireless networks,'' \emph{IEEE Communications Surveys \& Tutorials},
  vol.~14, no.~2, pp. 322--337, 2012.

\bibitem{Sadr2009}
S.~Sadr, A.~Anpalagan, and K.~Raahemifar, ``Radio resource allocation
  algorithms for the downlink of multiuser {OFDM} communication systems,''
  \emph{IEEE Communications Surveys \& Tutorials}, vol.~11, no.~3, pp. 92--106,
  2009.

\bibitem{Hoo2004}
L.~M. Hoo, B.~Halder, J.~Tellado, and J.~M. Cioffi, ``Multiuser transmit
  optimization for multicarrier broadcast channels: asymptotic {FDMA} capacity
  region and algorithms,'' \emph{IEEE Transactions on Communications}, vol.~52,
  no.~6, pp. 922--930, 2004.

\bibitem{Kim2005}
K.~Kim, Y.~Han, and S.-L. Kim, ``Joint subcarrier and power allocation in
  uplink {OFDMA} systems,'' \emph{IEEE Communications Letters}, vol.~9, no.~6,
  pp. 526--528, 2005.

\bibitem{Yu2006}
W.~Yu and J.~M. Cioffi, ``Constant-power waterfilling: performance bound and
  low-complexity implementation,'' \emph{IEEE Transactions on Communications},
  vol.~54, no.~1, pp. 23--28, 2006.

\bibitem{Yu2002}
W.~Yu, G.~Ginis, and J.~M. Cioffi, ``Distributed multiuser power control for
  digital subscriber lines,'' \emph{Selected Areas in Communications, IEEE
  Journal on}, vol.~20, no.~5, pp. 1105--1115., 2002.

\bibitem{Wong2008}
I.~C. Wong and B.~L. Evans, ``Optimal downlink {OFDMA} resource allocation with
  linear complexity to maximize ergodic rates,'' \emph{IEEE Transactions on
  Wireless Communications}, vol.~7, no.~3, pp. 962--971, 2008.

\bibitem{Jang2003}
J.~Jang and K.~B. Lee, ``Transmit power adaptation for multiuser {OFDM}
  systems,'' \emph{IEEE Journal on selected areas in communications}, vol.~21,
  no.~2, pp. 171--178, 2003.

\bibitem{Love2008}
D.~J. Love, R.~W. Heath, V.~K. Lau, D.~Gesbert, B.~D. Rao, and M.~Andrews, ``An
  overview of limited feedback in wireless communication systems,'' \emph{IEEE
  Journal on selected areas in Communications}, vol.~26, no.~8, pp. 1341--1365,
  2008.

\bibitem{Al-Harthi2007}
Y.~S. Al-Harthi, A.~H. Tewfik, and M.-S. Alouini, ``Multiuser diversity with
  quantized feedback,'' \emph{IEEE Transactions on Wireless Communications},
  vol.~6, no.~1, pp. 330--337, 2007.

\bibitem{Toufik2006}
I.~Toufik and H.~Kim, ``{MIMO-OFDMA} opportunistic beamforming with partial
  channel state information,'' in \emph{Communications, 2006. ICC'06. IEEE
  International Conference on}, vol.~12, 2006, pp. 5389--5394.

\bibitem{Svedman2007}
P.~Svedman, S.~K. Wilson, L.~J. Cimini, and B.~Ottersten, ``Opportunistic
  beamforming and scheduling for {OFDMA} systems,'' \emph{IEEE Transactions on
  Communications}, vol.~55, no.~5, pp. 941--952, 2007.

\bibitem{Naparstek2013}
O.~Naparstek and A.~Leshem, ``A fast matching algorithm for asymptotically
  optimal distributed channel assignment,'' in \emph{Digital Signal Processing
  (DSP), 2013 18th International Conference on}, 2013.

\bibitem{Naparstek2014}
------, ``Fully distributed optimal channel assignment for open spectrum
  access,'' \emph{IEEE Transactions on Signal Processing}, vol.~62, no.~2, pp.
  283--294, 2014.

\bibitem{Naparstek2014a}
------, ``Expected time complexity of the auction algorithm and the push
  relabel algorithm for maximum bipartite matching on random graphs,'' in
  \emph{Random Structures \& Algorithms}.\hskip 1em plus 0.5em minus
  0.4em\relax Wiley Online Library, 2014.

\bibitem{Bistritz2015}
I.~Bistritz and A.~Leshem, ``Asymptotically optimal distributed channel
  allocation: a competitive game-theoretic approach,'' in \emph{Communication,
  Control, and Computing (Allerton), 2015 53nd Annual Allerton Conference on},
  2015.

\bibitem{Bistritz2018}
------, ``Game theoretic dynamic channel allocation for frequency-selective
  interference channels,'' \emph{IEEE Transactions on Information Theory},
  2018, {DOI}: 10.1109/TIT.2018.2868440.

\bibitem{Papadimitriou98}
C.~H. Papadimitriou and K.~Steiglitz, \emph{Combinatorial optimization:
  algorithms and complexity}.\hskip 1em plus 0.5em minus 0.4em\relax Courier
  Corporation, 1998.

\bibitem{Lehmer1964}
D.~H. Lehmer, \emph{Applied Combinatorial Mathematics}, E.~F. Beckenbach,
  Ed.\hskip 1em plus 0.5em minus 0.4em\relax John Wiley \& Sons, 1964.

\bibitem{Siddique}
A.~B. Siddique, S.~Farid, and M.~Tahir, ``Proof of bijection for combinatorial
  number system,'' arXiv preprint arXiv:1601.05794. 2016.

\bibitem{LTE2009}
{LTE ETSI}, ``Evolved universal terrestrial radio access (e-utra); base station
  (bs) radio transmission and reception (3gpp ts 36.104 version 8.6. 0 release
  8), july 2009,'' \emph{ETSI TS}, vol. 136, no. 104, p.~V8, 2009.

\bibitem{Bistritz2017}
I.~Bistritz and A.~Leshem, ``Game theoretic resource allocation for m-dependent
  channel with application to {OFDMA},'' in \emph{Acoustics, Speech and Signal
  Processing (ICASSP), 2017 IEEE International Conference on}, 2017, pp.
  4316--4320.

\bibitem{Cover2012}
T.~M. Cover and J.~A. Thomas, \emph{Elements of information theory}.\hskip 1em
  plus 0.5em minus 0.4em\relax John Wiley \& Sons, 2012.

\bibitem{Rudin1964}
W.~Rudin \emph{et~al.}, \emph{Principles of mathematical analysis}.\hskip 1em
  plus 0.5em minus 0.4em\relax McGraw-hill New York, 1964.

\end{thebibliography}

\end{document}